\documentclass[twocolumn,superscriptaddress,longbibliography,aps,nofootinbib]{revtex4-2}
\usepackage{times}
\usepackage{color}
\usepackage{xspace}
\usepackage{amsmath}
\usepackage{amssymb}
\usepackage{amsbsy}
\usepackage{graphicx}
 \usepackage{bm}
 \usepackage{float}
\usepackage{relsize}
\usepackage[normalem]{ulem} 
\usepackage{lipsum}

\usepackage[unicode,breaklinks]{hyperref}
\hypersetup{
    unicode=true,
    plainpages=false, 
    colorlinks=true,
    linkcolor=blue,
    citecolor=blue,
    filecolor=black,
    urlcolor=blue
}

\newcommand{\bx}{\mathbf{x}}
\newcommand{\bk}{\mathbf{k}}

\newcommand{\Ut}{\tilde{U}}

\begin{document}

\title{Stability of a flattened dipolar binary condensate: emergence of the spin roton}
 
\author{Au-Chen Lee}
\affiliation{Department of Physics, Centre for Quantum Science,
and Dodd-Walls Centre for Photonic and Quantum Technologies, University of Otago, Dunedin, New Zealand}
\author{D. Baillie}
\affiliation{Department of Physics, Centre for Quantum Science,
and Dodd-Walls Centre for Photonic and Quantum Technologies, University of Otago, Dunedin, New Zealand}
\author{P. B. Blakie}
\affiliation{Department of Physics, Centre for Quantum Science,
and Dodd-Walls Centre for Photonic and Quantum Technologies, University of Otago, Dunedin, New Zealand}

\begin{abstract}
We develop theory for a two-component miscible dipolar condensate in a planar trap. Using numerical solutions and a variational theory we solve for the excitation spectrum and identify regimes where density- and spin-roton excitations are favored. We characterize the various instabilities that can emerge in this system over a wide parameter regime and present results for the stability phase diagram.   Importantly this allows us to identify the parameter regimes where a novel roton-immiscibility transition can occur, driven by the softening of the spin roton excitation.
\end{abstract}

\maketitle

\section{Introduction}

Two-component Bose-Einstein condensates in which one or both of the components have a large magnetic moment, present a new class of superfluids for exploring the interplay of mixture physics and long-ranged dipole-dipole interactions.  Fantastic recent experimental progress has led to a number of possible realizations. Most notably, the production of Bose-Einstein condensates of two different isotope mixtures of the highly magnetic atoms Er and Dy \cite{Trautmann2018,Politi2022a}, and the demonstration of magnetic Feshbach resonances to control the short ranged intra- and inter-species interactions of these mixtures \cite{Durastante2020}. Also by suppressing dipolar relaxation it may be  possible to realize multi-component superfluids using several different spin states of a single isotope  \cite{Chalopin2020}.  Another possibility involves a mixture of a highly magnetic atom with a weakly- or non-magnetic atom (cf.~related work on degenerate fermionic mixtures \cite{Ravensbergen2018a,Ravensbergen2020a}).
 The case of a two-component (binary) magnetic superfluid has been the subject of theoretical proposals for a new class of quantum droplet states \cite{Bisset2021,Smith2021a,Smith2021b} and supersolid phases \cite{Saito2009,Scheiermann2022a,Li2022a,Bland2022a}. The interplay of immiscibility and the long-ranged interactions has been the subject of investigations exploring pattern formation and novel instabilities \cite{Wilson2012a,KuiTian2018}. Also, a number of studies have considered properties of the condensate ground states in stationary \cite{Lee2021b} and in rotating frames \cite{Zhang2015,Zhang2016,Kumar2017,Shirley2014}  (also see  \cite{Pradas2022}).

\begin{figure}[htp!]   
 	\centering
 	\includegraphics[width=3.4in]{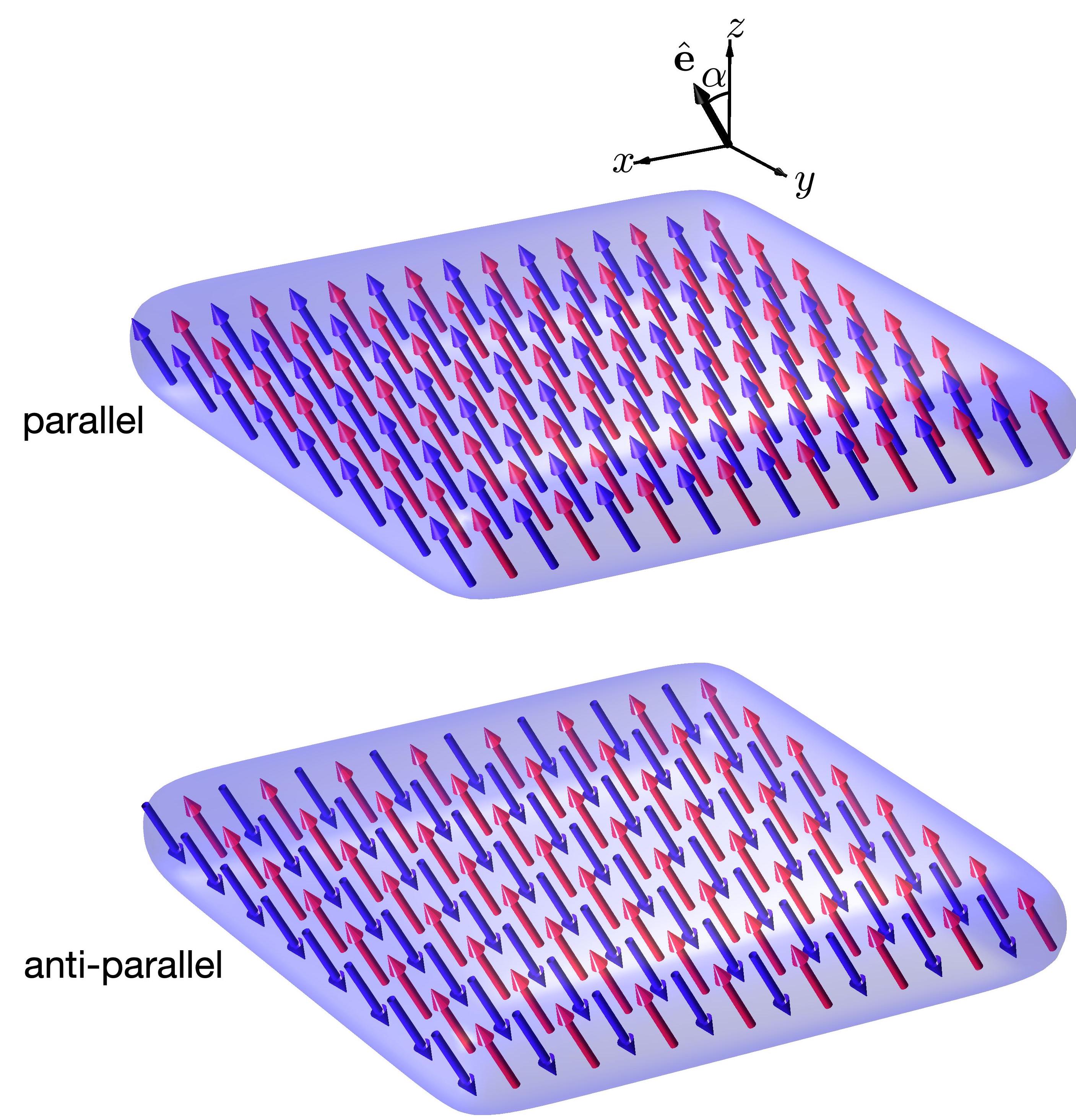}   
	 	\caption{Schematic showing a miscible binary dipolar BEC where the components have (upper) parallel and (lower) antiparallel polarization of dipole moments. The red and blue arrows represent the dipole moments of components 1 and 2, respectively.}\label{Fig:schematic}
 \end{figure}

The binary magnetic superfluid has a large number of parameters, including three contact interaction strengths, and the moments and orientation of the magnetic dipoles underlying the long-ranged dipole-dipole interactions (DDIs). This presents a rich landscape for exploring new behavior. 
In recent work the miscibility and stability of a dipolar condensate mixture was studied in a harmonic trap \cite{Lee2021b}. This work found that criteria derived from a homogeneous theory provided a good quantitative prediction of the instabilities for various regimes, with the general observation that the miscible regime was reduced as the strength of the DDIs increased. However, in pancake geometry traps (with the magnetic dipoles polarized along the tightly confined direction), the miscible regime extended over a much wider range of interaction parameters in quantitative disagreement with the homogeneous predictions. 
These results motivate the need for theoretical understanding of the physics of such a mixture in a flattened or planar trap.

Here we present theory for a binary dipolar condensate confined in a planar trap,  allowing for the dipole polarization of the components to vary, including the extreme cases of parallel and anti-parallel dipoles (see Fig.~\ref{Fig:schematic}). Our theory is developed for the case of the miscible phase where the two components overlap and are uniform in the plane of the trap. We solve for the quasiparticle excitation spectrum to study the collective modes of the system, and explore the conditions under which roton-like excitations emerge from the momentum dependence of the DDIs in the presence of confinement.  These rotons can be of density, (pseudo) spin density, or mixed character, as can be revealed from looking at the density or spin-density  structure factors. We use the excitation spectrum to locate the instabilities of the system, identified by excitations becoming dynamically unstable.  These instabilities can be categorised by various features, including whether the unstable excitation causes density or spin-density fluctuations, and whether it is a phonon (long wavelength) or roton (short wavelength). From this analysis we obtain stability phase diagrams, identifying where the various types of instabilities arise as a function of system parameters. We also develop various analytic estimates to identify the boundaries of these instabilities, thus providing a useful characterisation of the system over a wide parameter regime.

The outline of the paper is as follows. In Sec.~\ref{Sec:Form} we present the main formalism of the paper for the uniform miscible ground states and excitations. We also specialize the formalism to the balanced case, where the areal density and intra-species contact interaction of each component is taken to be the same.  This simplifies the treatment of the system and affords a variational approximation. 
Results for the balanced system are presented in Sec.~\ref{Sec:balancedresults}. Here our focus is on the stability phase diagrams for uniform miscible ground states and identifying the character of the instabilities that dominate in various parameter regimes. This analysis is aided by criteria developed from the limiting behavior of the interactions. We also consider how the roton or spin-rotons manifest in the density or spin-density structure factors.
In Sec.~\ref{Sec:unbalancedresults} we generalize our consideration and present results for the unbalanced case, before concluding in Sec.~\ref{Sec:conclusion}.

\section{Formalism}\label{Sec:Form}
Here we consider a planar binary Bose-Einstein condensate of magnetic atoms with axial harmonic confinement of angular frequency $\omega_z$ and no confinement in the transverse plane (see Fig.~\ref{Fig:schematic}). We develop our theory under the assumption that the two atomic components have the same mass $M$ and experience that same confinement potential. This should be a good approximation for a mixture of any two bosonic dysprosium or erbium isotopes, where the mass varies by less than 5\% and the ac-polarizability is similar (e.g.~see \cite{Trautmann2018}).

\subsection{\label{s:mftheory}Meanfield theory}
The meanfield theory for this system is provided by the two-component dipolar Gross-Pitaevskii equation (GPE) \cite{Goral2002}. Focusing on a miscible phase we take the component wavefunctions to be translationally invariant in the transverse plane, i.e.~$\Psi_i(\mathbf{x})=\sqrt{n_i}\psi_i(z)$, where $i=1,2$ is the component label, and $n_i$ is the areal component density with $\int dz |\psi_i|^2=1$. 

 The magnetic dipoles of each component are taken to be polarized along the axis $\hat{\mathbf{e}}=\hat{\mathbf{z}}\cos\alpha+\hat{\mathbf{x}}\sin\alpha$, that is at an angle of $0\le\alpha<\tfrac\pi2$ to the $z$-axis and lying in the $xz$-plane  (see Fig.~\ref{Fig:schematic}).  The DDI potential is of the form
 \begin{align}
 U^{dd}_{ij}(\mathbf{r})=\frac{3g^{dd}_{ij}}{4\pi}\frac{1-3\left(\hat{\mathbf{e}}\cdot\hat{\mathbf{r}}\right)^2}{r^3},
 \end{align}
where $g_{ij}^{dd}=\mu_0\mu_i^m\mu_j^m/3$ is the DDI coupling constant between atoms in components $i$ and $j$, with  $\mu_i^m$ being the magnetic moment of component $i$ on the $\hat{\mathbf{e}}$-axis. Note that we also consider the case of anti-parallel dipoles (see Fig.~\ref{Fig:schematic}) where  $\mu_1^m>0$ and $\mu_2^m<0$, for which $g_{12}^{dd}<0$ and $g_{ii}^{dd}>0$.

With these choices the  GPE for component $i$ takes the form
\begin{equation}
\mathcal{L}_i\psi_i(z) = \mu_i\psi_i(z) ,\label{tiGPE}
\end{equation}
where   $\mu_i$ is the associated chemical potential. The GPE operator is  given by
\begin{align}
\mathcal{L}_i= h_{z}+n_ig_{ii}|\psi_i|^2 +\sqrt{n_1n_2}g_{12}|\psi_{3-i}|^2, \label{Eq:GPEOp}
\end{align}
 where $h_{z} =-\frac{\hbar^2}{2M}\frac{d^2}{d{z}^2}+\frac{1}{2}M\omega_z^2{z}^2$, 
 \begin{align}
     g_{ij}=\lim_{k_\rho\to0} \Ut_{ij}(\bk) = g_{ij}^s+g_{ij}^{dd}(3\cos^2\alpha-1),\label{Eqgi}
 \end{align}
$g_{ij}^s=4\pi a_{ij}^s\hbar^2/M$ is the $s$-wave coupling constant describing the short ranged interactions between components $i$ and $j$, and we have introduced the interaction potential in $k$-space\footnote{The Fourier transform of the full interaction $U_{ij}(\mathbf{r})=g^s_{ij}\delta(\mathbf{r})+U^{dd}_{ij}(\mathbf{r})$}
\begin{align}
\Ut_{ij}(\mathbf{k})=g_{ij}^s+g_{ij}^{dd}\left[\frac{3\left(k_z\cos\alpha+k_x\sin\alpha\right)^2}{k^2}-1\right].\label{Eq:non_cutoff_V_K}
\end{align}

\subsection{\label{s:excitations}Excitations}
We can quantify the excitations of the condensate by solving for the Bogoliubov-de Gennes (BdG) quasi-particles.
The BdG equations can be obtained by linearizing the time-dependent GPE  $i\hbar\dot{\Psi}_j=\mathcal{L}_j\Psi_j$ about a stationary solution as
\begin{equation}
\Psi_j(\bx,t)=e^{-i\mu_j t}\left[\sqrt{n_j}\psi_j(z)+\vartheta_j(\bx,t)\right],
\end{equation}
where $\psi_j$ is a solution of the time-independent GPE (\ref{tiGPE}) and $\vartheta_j(\bx,t)$ is taken to be a small fluctuation of the form 
\begin{align}
&\vartheta_j(\bx,t)\equiv\label{vartheta}\\
&\sum_{\nu,\bk_\rho}\left[c_{\nu\bk_\rho} u_{\nu\bk_\rho j} e^{i\bk_\rho\cdot\bm{\rho}} e^{-i\omega_{\nu\bk_\rho} t}-c_{\nu\bk_\rho}^{*}v_{\nu\bk_\rho j}^{*} e^{-i\bk_\rho\cdot\bm{\rho}}e^{i\omega_{\nu\bk_\rho}^{*} t}\right].\nonumber
\end{align} 
Here $\bm{\rho}=(x,y)$ is the planar position vector,   $\nu$ labels the excitation band, and $\bk_\rho=(k_x,k_y)$ is the planar momentum vector.
The expansion (\ref{vartheta}) introduces the  quasi-particle modes $\mathbf{u}_{\nu\bk_\rho}(z)=(u_{\nu\bk_\rho1}(z),u_{\nu\bk_\rho2}(z))$ and $\mathbf{v}_{\nu\bk_\rho}(z)= ({v}_{\nu\bk_\rho1}(z),{v}_{\nu\bk_\rho2}(z))$, with excitation frequency $\omega_{\nu\bk_\rho}$, where  $c_{\nu\bk_\rho}$ is the (assumed small) amplitude of the mode. 

The  BdG equations that must be solved to determine the quasi-particles take the form\begin{align}
H_{k_\rho}\mathbf{w}_{\nu\bk_\rho}=\hbar\omega_{\nu\bk_\rho}\mathbf{w}_{\nu\bk_\rho},\label{BdG}
\end{align} 
where $\mathbf{w}_{\nu\bk_\rho}= (\mathbf{u}_{\nu\bk_\rho},  \mathbf{v}_{\nu\bk_\rho})^T$ and
\begin{equation}
H_{\bk_\rho}=
\begin{pmatrix}
\mathbf{L}_{\bk_\rho} +\mathbf{X}_{\bk_\rho}&-\mathbf{X}_{\bk_\rho}\\
\mathbf{X}_{\bk_\rho}&-\mathbf{L}_{\bk_\rho} -\mathbf{X}_{\bk_\rho}
\end{pmatrix},
\end{equation}
which contains two sub-matrix operators
\begin{equation}
\mathbf{L}_{\bk_\rho}=\begin{pmatrix}
\mathcal{L}_{1}+\epsilon_{k_\rho}-\mu_1&0\\
0&\mathcal{L}_{2}+\epsilon_{k_\rho}-\mu_2
\end{pmatrix},
\end{equation}
and 
\begin{equation}
\mathbf{X}_{\bk_\rho}=\begin{pmatrix}
X_{{\bk_\rho}11}&X_{{\bk_\rho}12}\\
X_{{\bk_\rho}21}&X_{{\bk_\rho}22}
\end{pmatrix}.
\end{equation}
Here $\epsilon_{k_\rho}=\hbar^2k_\rho^2/2M$ and  
\begin{align}
X_{{\bk_\rho}ij}f =\sqrt{n_in_j}\psi_i(z)\mathcal{F}_z^{-1}\!\left\{\Ut_{ij}(\bk_\rho+k_z\hat{\mathbf{z}})\mathcal{F}_z\!\left\{\psi_j(z)f \right\}\!\right\},
\end{align}
where $\mathcal{F}_z$ denotes the one-dimensional Fourier transform of $z$ coordinate, and $\mathcal{F}_z^{-1}$ denotes the inverse transform.
\subsection{General numerical treatment}
 For the numerical results presented in this paper we follow the methods shown in \cite{Baillie2015a} closely. Importantly, we  make use of a truncated interaction potential \cite{Ronen2006a,Baillie2015a} to reduce finite grid range effects in the evaluation of the DDIs.  For other aspects related to solving for the ground state and excitations we use the techniques described in \cite{Lee2021a,Lee2021b}. 
 
\subsection{Spin and density structure factors}\label{Sec:DSF}
The dynamic structure factor is a measure of the density fluctuations in a system, and can be used to characterize its structure and collective excitations.
For a two component system the dynamic structure factor can be generalized to characterize the \textit{density} ($+$) or \textit{spin-density} ($-$) fluctuations in the plane and is defined as (see Ref.~\cite{Lee2021b})
\begin{align}
	S_\pm(\bk_\rho,\omega)&\equiv\sum_{\nu}|\delta n^{\pm}_{\nu\bk_\rho}|^2\delta(\omega-\omega_{\nu\bk_\rho}),\label{Eq.SF}
\end{align}
where $\delta n^{\pm}_{\nu\bk_\rho}\equiv\delta n_{\nu\bk_\rho1}\pm\delta n_{\nu\bk_\rho2}$,
with
\begin{align}
	\delta n_{\nu\bk_\rho j}&=\int dz\,{\psi}_j(z)\left[{u}_{\nu\bk_\rho j}(z)-{v}_{\nu\bk_\rho j}(z)\right],
\end{align}
being the integral of the density fluctuation of component $j$ over $z$. 
The density dynamic structure factor has been measured in dipolar quantum gases using Bragg spectroscopy (e.g.~see \cite{Petter2019,Petter2021a}). The dynamic structure factor, and the static structure factor defined by 
 \begin{align}
 S_\pm(\bk_\rho)\equiv\int d\omega\,S_\pm(\bk_\rho,\omega),
 \end{align}
are particularly sensitive probes to roton excitations which we consider in the results sections.	 
	 
\subsection{Balanced regime}\label{Sec:BalancedTheory}
In this work we initially focus on results for the \textit{balanced regime} where the component densities and intra-species interactions are identical, i.e.~$n=n_1=n_2$,  $g_{11}^s=g_{22}^s$, and $|\mu_1^m|=|\mu_2^m|$. This latter condition means that $g^{dd}_{11}=g^{dd}_{22}=|g^{dd}_{12}|$, but allows for two possible orientation classes:  \textit{parallel dipoles}  (i.e.~$\mu_1^m=\mu_2^m$) with $g^{dd}_{12}=g^{dd}_{ii}$, which we denote as $\uparrow\uparrow$; and   \textit{anti-parallel dipoles} (i.e.~$\mu_1^m=-\mu_2^m$) with $g^{dd}_{12}=-g^{dd}_{ii}$, which we denote as $\uparrow\downarrow$.

The balanced configuration allows us to concentrate on how the excitations of the system depend on the effect of the interspecies contact interaction $g^s_{12}$, the relative strength of the DDIs to the contact interactions, and the parallel versus anti-parallel orientation of the dipoles.

Under the balanced conditions, when the system is miscible, the ground state profile is the same for both components, we set $\psi_j\to\psi$ with $\mu_j\to\mu$. The GPE (\ref{tiGPE}) reduces to
\begin{align}
\mu\psi&=\mathcal{L}\psi,\\
\mathcal{L}&= h_{z}+ng_{0+}^p|\psi|^2,
\label{GPEopbalanced}
\end{align}
where $g_{0+}^p$ denotes the effective condensate interaction with $p=\{\uparrow\uparrow,\uparrow\downarrow\}$ denoting the two dipole orientations:
\begin{align}
g_{0+}^{\uparrow\uparrow}&=%
g_{ii}^s+g_{12}^s+2g_{ii}^{dd}(3\cos^2\alpha-1),\label{g+uu}\\
g_{0+}^{\uparrow\downarrow}&= g_{ii}^s+g_{12}^s.\label{g+ud}
\end{align}  
 Notably, the effect of the DDIs on the ground state cancel for the anti-parallel case.

For the balanced case the excitations can be further chosen to have the components fluctuating in-phase ($\lambda=+1$) or out-of-phase ($\lambda=-1$), i.e.
\begin{align}
\sigma_x\mathbf{u}_{\lambda \nu\bk_\rho}=\lambda \mathbf{u}_{\lambda \nu\bk_\rho},\quad \sigma_x\mathbf{v}_{\lambda \nu\bk_\rho}=\lambda \mathbf{v}_{\lambda \nu\bk_\rho},\quad
\end{align}
where  $\sigma_x= \left[\begin{smallmatrix} 0 & 1 \\  1 & 0 \end{smallmatrix} \right] $ is the $x$ Pauli matrix, i.e.~ $\mathbf{u}_{\pm \nu\bk_\rho}=(u_{\pm\nu\bk_\rho}(z),\pm u_{\pm\nu\bk_\rho}(z))$, etc.
 The BdG equations are then block diagonal and can be decoupled into the two uncoupled 2-by-2 systems:
 \begin{align} 
H_{\lambda}\begin{pmatrix} u_{\lambda \nu \bk_\rho} \\ v_{\lambda \nu \bk_\rho}\end{pmatrix}=\hbar\omega_{\lambda \nu \bk_\rho} \begin{pmatrix} u_{\lambda \nu \bk_\rho} \\ v_{\lambda \nu \bk_\rho}\end{pmatrix},
 \end{align}
 where 
 \begin{align}
H_{\lambda}&=\begin{pmatrix}
\mathcal{L}+\epsilon_{k_\rho} +X_{\lambda}-\mu&-X_{\lambda} \\
X_{\lambda}&-(\mathcal{L}+\epsilon_{k_\rho} +X_{\lambda}-\mu )
\end{pmatrix},\label{e:Hlambda}\\
X_{\lambda}f &=n\psi(z)\mathcal{F}_z^{-1}\!\left\{ \Ut_{\lambda}(\bk_\rho+k_z\hat{\mathbf{z}})\mathcal{F}_z\!\left\{\psi(z)f \right\}\!\right\},\label{Xf}
\end{align}
 where we define
 \begin{align}
     \Ut_{\lambda}(\bk)=\sqrt{\Ut_{11}(\bk)\Ut_{22}(\bk)}+\lambda \Ut_{12}(\bk),\label{Ulambda}
 \end{align}
 noting that in the balanced case $\Ut_{\lambda}(\bk)=\Ut_{ii}(\bk)+\lambda \Ut_{12}(\bk)$, and we discuss the extension to the unbalanced case in Sec.~\ref{s:approxstabcrit}.
 
 The excitations with $\lambda=+1$ describe (total) density fluctuations whereas those with $\lambda=-1$ describe (pseudo) spin fluctuations (also see discussion of the dynamic structure factors in Sec.~\ref{Sec:DSF}). For this reason we will refer to the relevant excitations as density and spin branches, and $\Ut_{+}$ ($\Ut_{-}$) as the density (spin) interaction.
 
  \begin{table*}[htb!]
\caption{\label{tab:Utilde} The characteristic properties of the interactions. }
     \begin{tabular}{l|c|c|c|c} \hline
      \multicolumn{5}{c}{balanced case} \\\hline
$p$ & \multicolumn{2}{c|}{$\uparrow\uparrow$}& \multicolumn{2}{c}{$\uparrow\downarrow$} \\ [0.5ex] 
\hline
$\lambda$ & $+$ & $-$& $+$ & $-$\\ [0.5ex] 
\hline
$g_{0\lambda}^p=\displaystyle\lim_{k_\rho\to0}\Ut_\lambda$  &  $g_{ii}^s+g_{12}^s+2g_{ii}^{dd}(3\cos^2\alpha-1)$ &  $g_{ii}^s-g_{12}^s$ &  $g_{ii}^s+g_{12}^s$ & $g_{ii}^s-g_{12}^s+2g_{ii}^{dd}(3\cos^2\alpha-1)$ \\ [0.5ex] 
 $g_{\Delta\lambda}^p\quad$ [see Eq.~(\ref{EqgDelta})] & $2g_{ii}^{dd}\cos^2\alpha$ & $0$ & $0$ & $2g_{ii}^{dd}\cos^2\alpha$ \\ [0.5ex] 
$g_{y\lambda}^p=\displaystyle\lim_{k_y\to\infty}\Ut_\lambda$ & $g^s_{ii}+g^s_{12}-2g_{ii}^{dd}$ & $g^s_{ii}-g^s_{12}$ & $g^s_{ii}+g^s_{12}$ & $g^s_{ii}-g^s_{12}-2g_{ii}^{dd}$ \\  [0.5ex] 
\hline
       \multicolumn{5}{c}{general results} \\\hline
       $g_{ij}=\displaystyle\lim_{k_\rho\to0}\Ut_{ij}$  &  \multicolumn{4}{l}{   $g^s_{ij}+g^{dd}_{ij}(3\cos^2\alpha-1)$}\\   
       $ g_{0\lambda}=\displaystyle\lim_{k_\rho\to0}\Ut_\lambda  $  &  \multicolumn{4}{l}{   $\sqrt{g_{11}g_{22}}+\lambda g_{12}$}\\
       $ g_{y\lambda}=\displaystyle\lim_{k_y\to\infty}\Ut_\lambda $  &  \multicolumn{4}{l}{   $\sqrt{(g^s_{11}-g^{dd}_{11})(g^s_{22}-g^{dd}_{22})} +\lambda (g^s_{12}-g^{dd}_{12})$}\\\hline
   
     \end{tabular}
     \end{table*}

\subsection{Properties of $\Ut_{\lambda}$}
 Many of the excitation properties we examine can be understood from the limiting behavior of $\Ut_{\lambda}(\bk)$. In this subsection we collect the key results that we use later.
 
 The long wavelength behavior is characterized by
 \begin{align}
g_{0\lambda}^p\equiv  \lim_{k_\rho\to0}\Ut_{\lambda}(\mathbf{k}).
 \end{align}
 This quantity is independent of $k_z$  [see~Eqs.~(\ref{g+uu}) and (\ref{g+ud})], and $g_{0+}^p$ corresponds to the condensate interaction. The full set of values for $g_{0\lambda}^p$, including those for $\lambda=-1$,
are given in Table \ref{tab:Utilde}.

It is convenient to express $\Ut_{\lambda}$ in the form\footnote{Noting that $\lim_{k_\rho\to0}\Ut_{\lambda}(\mathbf{k})=g_{0\lambda}^p$, $\lim_{k_y\to\infty}\Ut_{\lambda}(\mathbf{k})=g_{0\lambda}^p-3g^p_{\Delta \lambda}$, and $\lim_{k_x\to\infty}\Ut_{\lambda}(\mathbf{k})=g_{0\lambda}^p+3g^p_{\Delta \lambda}(\tan^2\alpha-1)$.} 
 \begin{align}
\Ut_{\lambda}(\mathbf{k})=g_{0\lambda}^p+3g^p_{\Delta \lambda}\left[\frac{\left(k_z +k_x\tan\alpha\right)^2}{k^2}-1 \right].
 \end{align}
With this identification we have introduced  
 \begin{align}
 g^p_{\Delta \lambda}\equiv  (g^{dd}_{ii}+\lambda g_{12}^{dd})\cos^2\alpha,\label{EqgDelta}
 \end{align} 
(also listed in Table \ref{tab:Utilde}).
In cases where $g^p_{\Delta \lambda}=0$ (i.e.~the density interaction for the anti-parallel case, and the spin interaction  for the parallel case), the interactions are momentum independent, and thus contact like. For cases where $g^p_{\Delta \lambda}\ne0$ we have to carefully look at the finite momentum excitations for the development of roton-like excitations.

It is also useful to consider the interactions for short-wavelength in-plane excitations. In practice for our choice of tilting dipoles in the $xz$-plane the most stringent conditions for stability (i.e.~least repulsive interactions) occurs along the $k_y$-axis, and thus we define
\begin{align}
g_{y\lambda}^p&\equiv \lim_{k_y\to\infty}\Ut_{\lambda}(\mathbf{k})= g_{0\lambda}^p-3g_{\Delta\lambda}^p,
\label{parallelUinflims-}
\end{align}
so that $g^p_{\Delta \lambda}= (g^p_{0 \lambda} - g^p_{y \lambda})/3$, with values for the particular cases evaluated in Table \ref{tab:Utilde}.

\subsection{Variational treatment}
A variational Gaussian treatment can be applied to the balanced case. In this approach each component wavefunction has the form 
\begin{align}
	\psi_{\sigma }(z)=\frac{1}{\pi^{1/4}\sqrt{\sigma l_z}}e^{-z^2/2{\sigma^2l_z^2}},
\end{align}
where $\sigma$ is the dimensionless variational axial width parameter and $l_z=\sqrt{\hbar/M\omega_z}$. The total energy of the condensate is given by
\begin{align}
	\frac{E(\sigma)}{\hbar\omega_z}=\frac{1}{2\sigma^2}+\frac{\sigma^2}{2}+\gamma\frac{ng_{0+}^p}{\hbar\omega_z},\label{Eq:variational_GPE}
\end{align}
where   $\gamma\equiv \int dz\,l_z|\psi_\sigma|^4=1/\sqrt{2\pi}\sigma$.

We can use the variational approximation to describe the lowest in-phase and the lowest out-of-phase excitation bands (i.e.~$\nu=0$ and $\lambda=\pm1$) by employing the ``same-shape approximation" \cite{Baillie2015a,Blakie2020a,Pal2020a}, i.e.~taking that  $u_{\lambda0\bk_\rho j} (z)\to \mathrm{u}_{\lambda\bk_\rho j}\psi_\sigma(z) $ and $v_{\lambda0\bk_\rho j} (z)\to  \mathrm{v}_{\lambda\bk_\rho j}\psi_\sigma(z)$, where $\{\mathrm{u}_{\lambda\bk_\rho j},\mathrm{v}_{\lambda\bk_\rho j}\}$ are constants. We can then integrate out the $z$ coordinates to obtain 
\begin{align}
	\hbar\omega_{\lambda\bk_\rho}\!=\!\sqrt{\epsilon_{k_\rho}\bigl\{\epsilon_{k_\rho}+2n\gamma \left[g_{0\lambda}^p+g_{\Delta\lambda}^pF(\mathbf{k}_\rho)\right]\bigr\}},\label{Eq:twocomp_in_singlecomp_form}
\end{align} 
where [cf.~Eq.~(\ref{Xf})]
\begin{align}
\!F(\mathbf{k}_\rho)\!&=\!\frac{3l_z}{\gamma}\!\!\int dz\psi_\sigma^2 \mathcal{F}_z^{-1}\!\biggl\{\!\biggl[\frac{\left(k_z +k_x\tan\alpha\right)^2}{k^2}-1 \biggr]\mathcal{F}_z\{\psi_\sigma^2  \}\!\biggr\},\\
    &=3 \left(\frac{k_x^2}{k_\rho^2}\tan^2\alpha-1\right)G_0(k_\rho \sigma l_z/\sqrt{2}), \label{Eq:Fij} 
	 \end{align}
with $G_0(q) = \sqrt\pi q e^{q^2}\mathrm{erfc}(q)$.

\section{Balanced system results}\label{Sec:balancedresults}

\subsection{Instabilities for the parallel case}\label{Sec:instabilitiesParallel}

 \begin{figure}[htp!] 
 \centering
 \includegraphics[width=\linewidth]{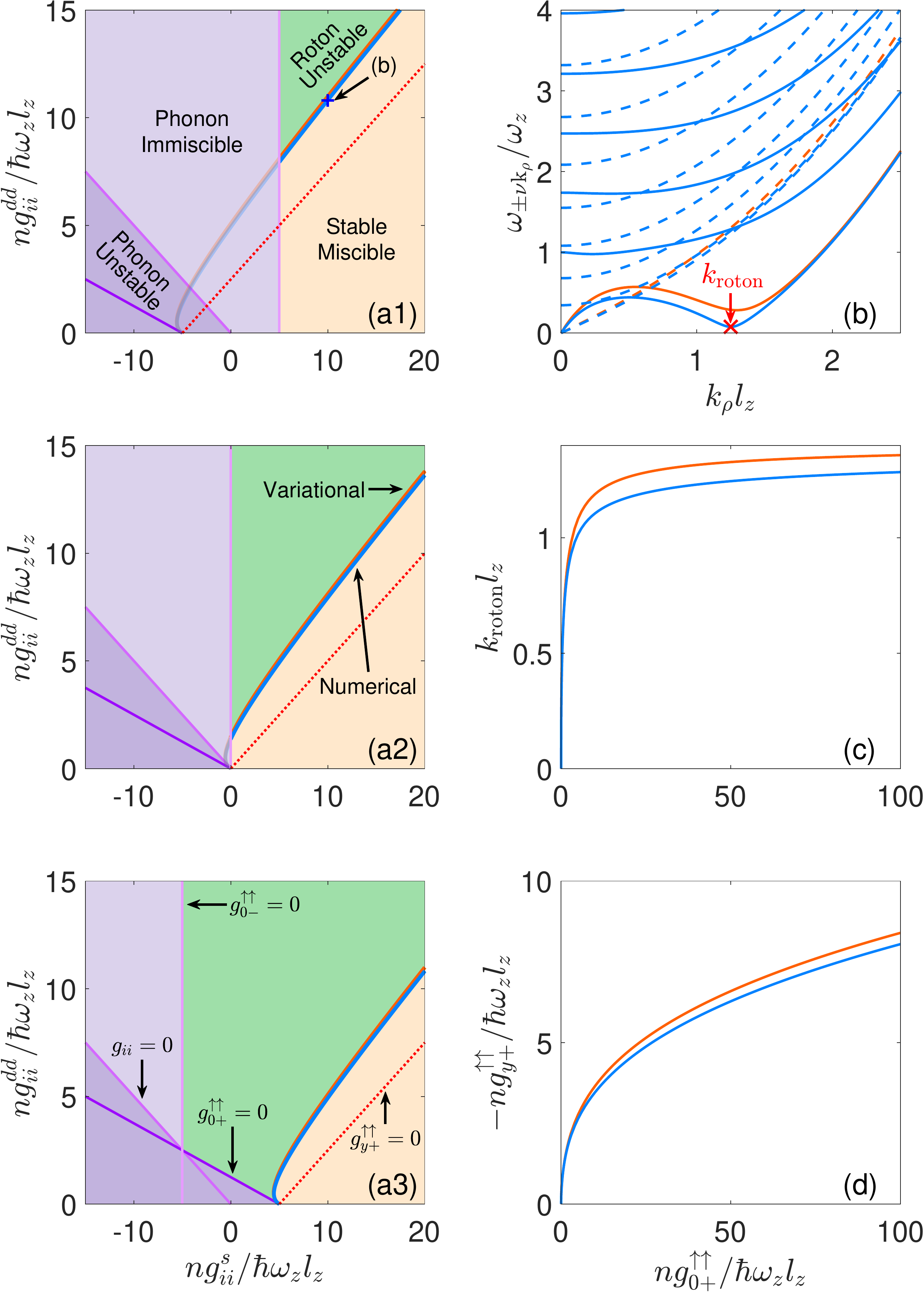}\\
 	\caption{(a1)-(a3) Stability phase diagram for a uniform balanced system with parallel dipoles. Results for $\alpha=0$ and (a1) $ng_{12}^s/\hbar\omega_zl_z=5$, (a2) $0$, and (a3) $-5$. The shaded regions identify where the uniform state is either stable as a miscible mixture, or has various instabilities as described in the main text.
    (b) Shows the excitation spectrum for $n(g_{ii}^s,g_{ii}^{dd})/\hbar\omega_zl_z = (10,10.8)$ [shown with a `$+$' symbol in subplot (a1)]. The  density $\lambda=+1$ (spin $\lambda=-1$) excitation branches are shown as solid (dashed) lines.  The roton minimum is indicated.
	(c) The roton wavevector at the point that the roton excitation softens to zero energy. 
	(d) The interaction parameters for the roton boundary.
	In subplots  (a)-(d) blues lines indicate the results from the full numerical calculation of the GPE and BdG equations, whereas the orange lines are from the variational theory.
	 }\label{Fig:PDparallel}
 \end{figure}

First we consider the instabilities that can occur for the case of a balanced binary dipolar fluid with parallel dipoles.  We show three example stability phase diagrams for this system in Figs.~\ref{Fig:PDparallel}(a1)-(a3). The various instabilities are identified from the type of excitation that becomes dynamically unstable (i.e.~the excitation develops an imaginary energy). 

\subsubsection{Density instabilities and the density roton}
The phonon (long-wavelength) interaction for this system, $g_{0+}^{\uparrow\uparrow}$, must be non-negative for the relevant modes of the system to be stable. We identify $g_{0+}^{\uparrow\uparrow}=0$ as the \textit{phonon instability boundary} (also see discussion of component stability below). 

Because $g_{\Delta+}^{\uparrow\uparrow}\ne0$ the density interaction is momentum dependent for the parallel system. If $g_{y+}^{\uparrow\uparrow}<0$  then the high-$k_y$ interaction is attractive and this can allow a roton excitation to form [e.g.~see Fig.~\ref{Fig:PDparallel}(b)]. For $g_{y+}^{\uparrow\uparrow}$ sufficiently attractive the roton excitation goes soft, marking the onset of a roton instability. The condition $g_{y+}^{\uparrow\uparrow}=0$ thus provides a lower bound for where a roton can form. In practice $g_{y+}^{\uparrow\uparrow}$ has to be sufficiently attractive for the interactions to overcome the kinetic energy of the excitation, allowing a local minimum to form in the dispersion. Thus, the precise conditions for where the rotons form have to be determined from a calculation of the excitation spectrum. Such a spectrum is shown in Fig.~\ref{Fig:PDparallel}(b)   for a case close to where the roton goes to zero energy. The wavevector of the roton $ k_\text{roton}$ is identified as the location of the local minimum in the lowest density-branch excitation (i.e.~$\omega_{\mathbf{k}_\rho0+}$). We numerically determine the roton boundary by searching for the interaction parameters where the roton energy is zero. The blue and orange lines in Figs.~\ref{Fig:PDparallel}(a1)-(a3) show the \textit{roton instability boundary} from the BdG solution and variational theory, respectively. Figure~\ref{Fig:PDparallel}(c)  shows the roton wavevector along the roton instability boundary and Fig.~\ref{Fig:PDparallel}(d) gives the interaction parameter coordinates of the roton instability boundary. 
The groundstate depends only on $g_{0+}^{\uparrow\uparrow}$, and the matrix $H_+$ \eqref{e:Hlambda} depends only on $g_{0+}^{\uparrow\uparrow}$ and $g_{y+}^{\uparrow\uparrow}$, so the density roton boundary $g_{y+}^{\uparrow\uparrow}$, can be written as a universal function of only $g_{0+}^{\uparrow\uparrow}$. 
We also note that comparison of numerical calculations of the  BdG equations to the variational results in these figures shows good qualitative agreement. 

The density modes of the parallel balanced system map onto those of an equivalent scalar dipolar gas (in the same planar geometry) with the identification of $g^s_{ii}+g^{s}_{12}$ and $2g^{dd}_{ii}$ as the effective contact and dipolar coupling contacts of the scalar system, respectively. The basic density instabilities of this system are thus identical to those of the scalar system explored in Ref.~\cite{Baillie2015a}. Importantly, the roton stability boundary is universal in the units of Fig.~\ref{Fig:PDparallel}(d) (c.f~Fig.~5(a) of Ref.~\cite{Baillie2015a}), and is the same curve appropriately mapped with the choice of $\alpha$ and differing $g_{12}^s$ values in Figs.~\ref{Fig:PDparallel}(a1)-(a3). Of course the additional features of the binary dipolar gas contained in the spin structure and the individual components (discussed below), add additional structure to the stability phase diagram.

\subsubsection{Spin instabilities}
The spin instabilities can be analyzed in a similar way to the density instabilities, but using the relevant results from the $\Ut_-$ interaction. However, as $g_{\Delta-}^{\uparrow\uparrow}=0$, there is no momentum dependent behavior in the spin interaction. Thus we only need to consider the long wavelength behavior characterized by  $g_{0-}^{\uparrow\uparrow}$. The condition $ g_{0-}^{\uparrow\uparrow}=0$ marks the boundary to a long wavelength instability. Because this occurs in the spin interaction, the instability marks the development of immiscibility (phase separation) of the two components, and we refer to this instability as \textit{phonon immiscibility}.

\subsubsection{Component stability}\label{Sec:SCSB}
For the uniform binary system to be stable, we also require that each individual component is stable. This requires that the coefficient of $|\psi_i|^2$ in Eq.~(\ref{Eq:GPEOp}) is non-negative. This coefficient corresponds to the long-wavelength limit $g_{ii}$. Thus we identify the condition for the component-$i$ stability boundary  as $g_{ii}=0$. 
 This marks the onset of a phonon instability in this component.
  For the results shown Figs.~\ref{Fig:PDparallel}(a1)-(a3) we see that the density and single component stability both contribute to determining the region in which the system has a phonon instability.

The density instabilities (of the phonon or roton form) and the single component instabilities, indicate a mechanical instability of the system where the total density or the density in a component can collapse and form density spikes. In this regime the quantum fluctuation effects are necessary to stabilize the collapse (e.g.~see \cite{Bisset2021,Smith2021a,Smith2021b}).

\subsection{Instabilities for the anti-parallel case}
\begin{figure}[htp!] 
 \centering
 \includegraphics[width=\linewidth]{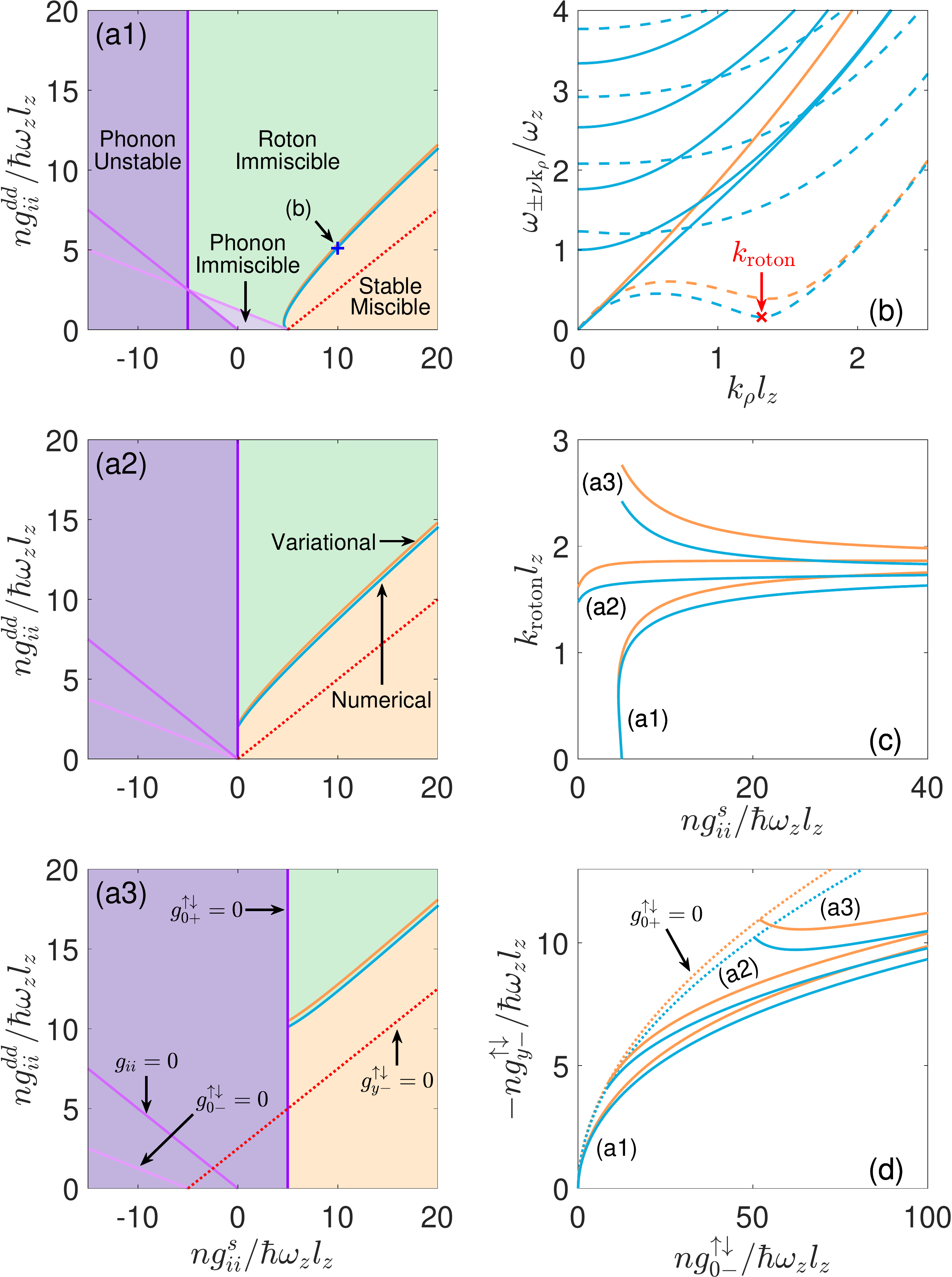}\\
 	\caption{(a1)-(a3) Stability phase diagram for a uniform balanced system with anti-parallel dipoles. Results for $\alpha=0$ and (a1) $ng_{12}^s/\hbar\omega_zl_z=5$, (a2) $0$, and (a3) $-5$. The shaded regions identify where the uniform state is either stable as a miscible mixture, or has various instabilities as described in the main text.
	(b) Shows the excitation spectrum for $n(g_{ii}^s,g_{ii}^{dd})/\hbar\omega_zl_z = (10,5.1)$ [shown with a `$+$' symbol in subplot (a1)]. The  density $\lambda=+1$ (spin $\lambda=-1$) excitation branches are shown as solid (dashed) lines. The roton minimum is indicated.
	(c) The roton wavevector at the point that the roton excitation softens to zero energy. 
 (d) The interaction parameters for the roton boundary.
	In subplots  (a)-(d) blues lines indicate the results from the full numerical calculation of the GPE and BdG equations, whereas the  orange lines are from the variational theory.
	 }\label{Fig:PDantiparallel}
 \end{figure}

We now consider the instabilities that can occur in a balanced binary dipolar fluid where the dipole moments of the two components are anti-parallel.  We show three example stability phase diagrams for this system in Figs.~\ref{Fig:PDantiparallel}(a1)-(a3). 

\subsubsection{Density instabilities}
The phonon interaction for this system is $g_{0+}^{\uparrow\downarrow}$ and we identify $g_{0+}^{\uparrow\downarrow}=0$ as a phonon instability boundary.   
Because $g_{\Delta+}^{\uparrow\downarrow}=0$ the density interaction is momentum independent and a (density) roton cannot occur.

\subsubsection{Spin instabilities and the spin-roton} 
The condition $g_{0-}^{\uparrow\downarrow}=0$ marks the long-wavelength phonon immiscibility transition.
Because $g_{\Delta-}^{\uparrow\downarrow}\ne0$ the spin interaction is momentum dependent for the anti-parallel system. 
If $g_{y-}^{\uparrow\downarrow}<0$  then a spin-roton excitation can form [e.g.~see Fig.~\ref{Fig:PDantiparallel}(b)]. The spin-roton going to zero energy marks the onset of a \textit{roton immiscibility transition}, i.e.~where phase separation is driven by an unstable mode of a finite wavevector $k_\text{roton}$.
The condition $g_{y-}^{\uparrow\downarrow}=0$ thus provides a lower bound for where spin-roton can form.   Figure~\ref{Fig:PDantiparallel}(c) shows the roton wavevector at the roton instability boundary and Fig.~\ref{Fig:PDantiparallel}(d) gives the interaction parameter coordinates of the roton immiscibility  boundary.  
  
We note that for the anti-parallel case, $H_-$ depends on $g_{0+}^{\uparrow\downarrow}$, $g_{0-}^{\uparrow\downarrow}$, and $g_{y-}^{\uparrow\downarrow}$, and the spin roton boundary $g_{y-}^{\uparrow\downarrow}$ is a function of both $g_{0\pm}^{\uparrow\downarrow}$. So, the roton immiscibility boundary does not display the universality found for the density roton in the parallel case [cf.~Fig.~\ref{Fig:PDparallel}(d)]. The critical $g_{y-}^{\uparrow\downarrow}$ line terminates when $g_{0+}^{\uparrow\downarrow}=0$ and a phonon mode becomes unstable.

\subsubsection{Component stability}
The single component stability condition for the anti-parallel system is $g_{ii}\ge0$, i.e. the same as for the parallel balanced system, as was given in Sec.~\ref{Sec:SCSB}.

\subsection{Structure factors}
\begin{figure}[htp!] 
	\centering
    \includegraphics[width=\linewidth]{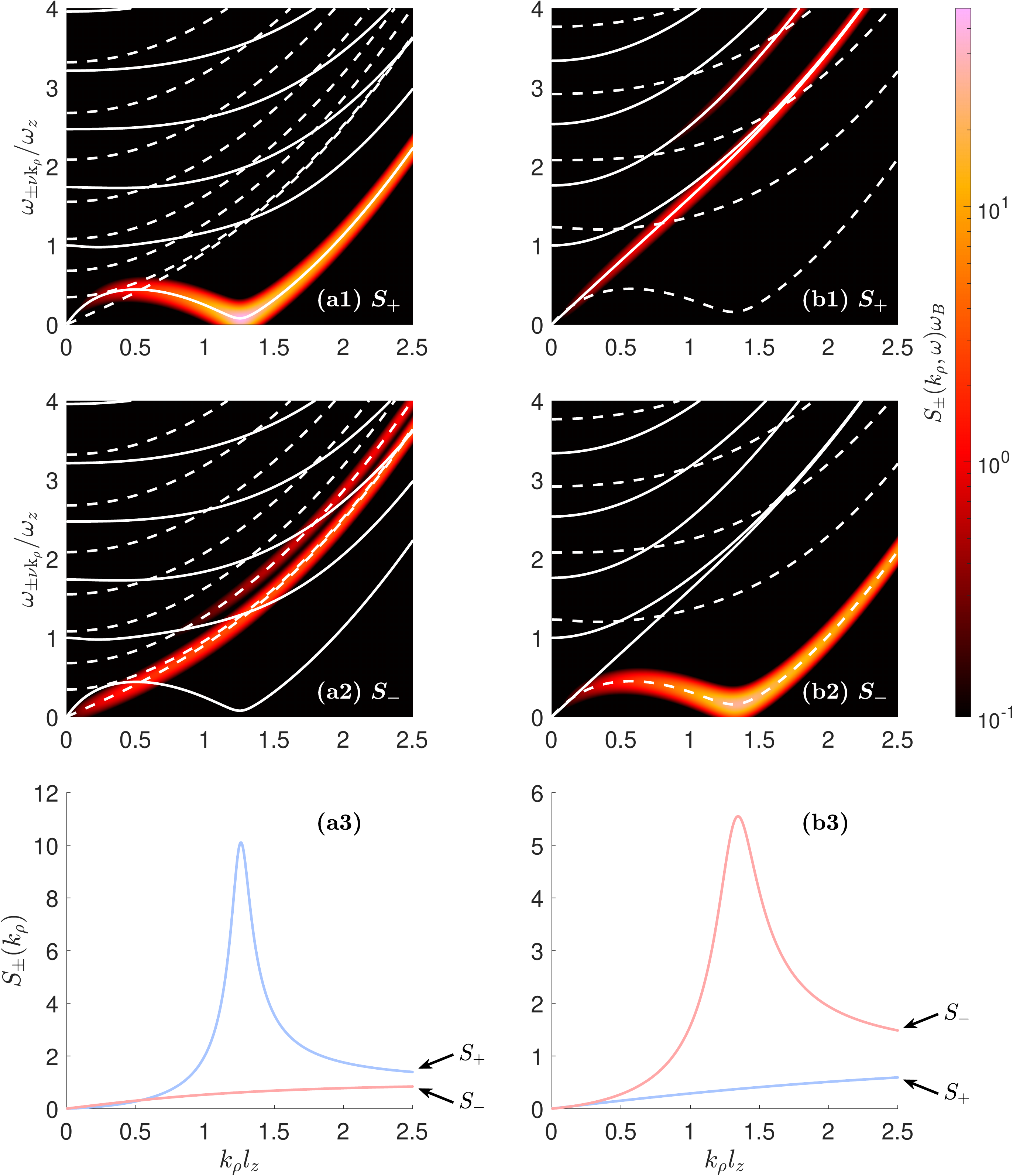}
	\caption{The density  and spin-density dynamic and static structure factors for a balanced system with parallel (a1)-(a3) and anti-parallel (b1)-(b3) dipole orientations.   Results in (a1)-(a3) are for the parallel case considered in Fig.~\ref{Fig:PDparallel}(b) and  (b1)-(b3) are for the anti-parallel case in Fig.~\ref{Fig:PDantiparallel}(b). White solid (dashed) lines show the in-phase/density (out-of-phase/spin) excitation spectrum for reference. The dynamic structure factors are frequency broadened by a frequency width of $\omega_B=0.1\omega_z$.     \label{Fig:DSF}}
\end{figure}
 
 In Fig.~\ref{Fig:DSF} we show the dynamic and static structure factors for the parallel and anti-parallel balanced cases. 
 
 The parallel case is shown in Figs.~\ref{Fig:DSF}(a1)-(a3) and corresponds to the spectrum shown in Fig.~\ref{Fig:PDparallel}(b), where we observed a density roton feature. The roton portion of the lowest $\lambda=+1$ excitation branch makes a strongly weighted contribution to the density dynamic structure factor [Fig.~\ref{Fig:DSF}(a1)] and appears as a prominent peak in the density static structure factor   [Fig.~\ref{Fig:DSF}(a3)].
 
The anti-parallel case [corresponding to the spectrum in Fig.~\ref{Fig:PDantiparallel}(b)] exhibits a spin-roton. This feature makes a strongly weighted contribution to the spin-density dynamic structure factor [Fig.~\ref{Fig:DSF}(b2)] and appears as a peak in the spin-density static structure factor  [Fig.~\ref{Fig:DSF}(b3)].

\section{Unbalanced system results} \label{Sec:unbalancedresults}

\subsection{Excitations in the unbalanced regime}

  \begin{figure}[htp!] 
	\centering
	\includegraphics[width=\linewidth]{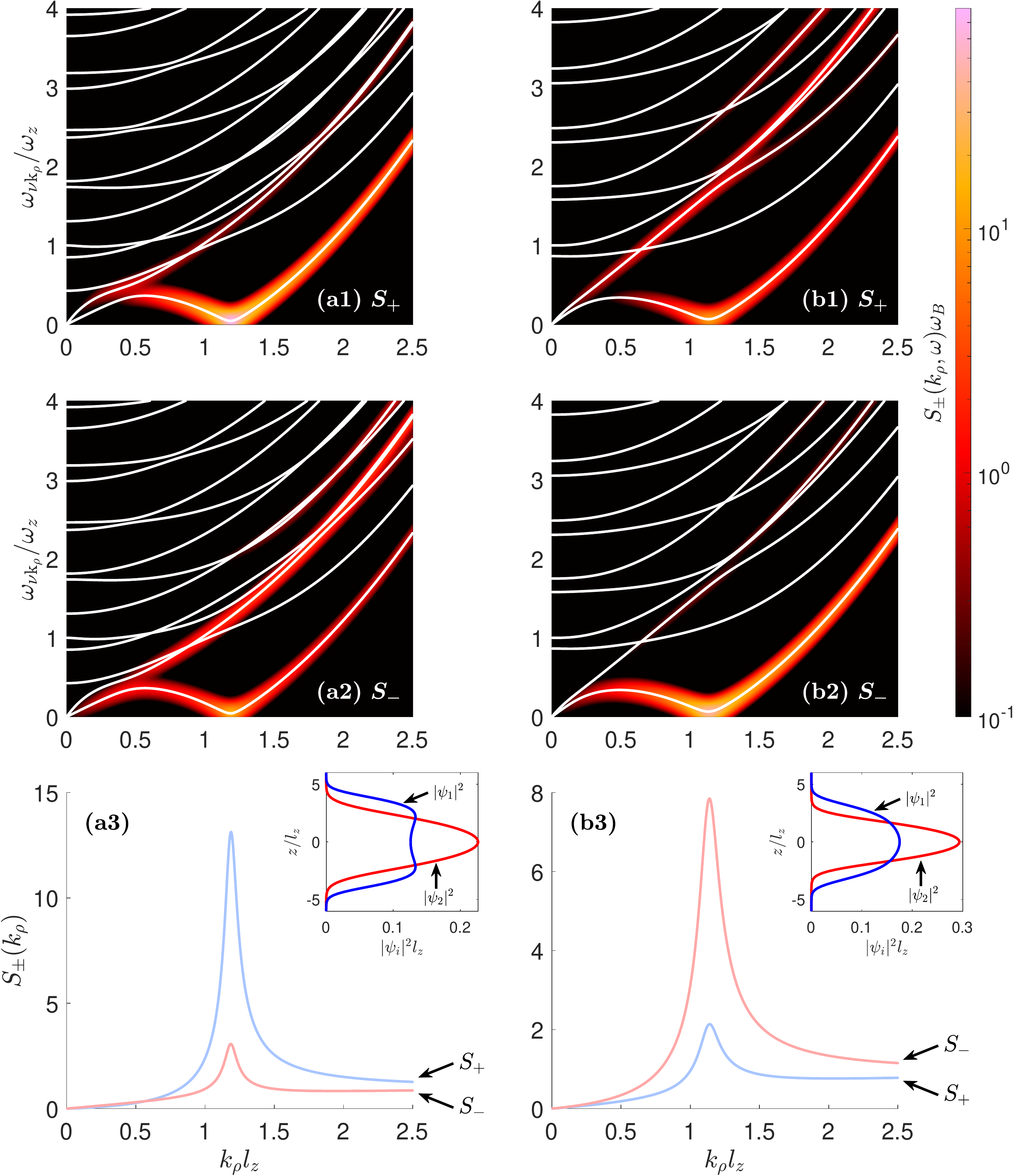}
	\caption{ The density  and spin-density dynamic and static structure factors for an unbalanced system with  $n(g^{dd}_{11}, g^{dd}_{22})/\hbar\omega_zl_z = (14,7)$ (a1)-(a3), $(13.4,0)$ (b1)-(b3), and other parameters as in Fig.~\ref{Fig:PDparallel}(b).  White solid lines show the excitation spectrum for reference. The insets to (a3) and (b3) show the axial density profiles $|\psi_1|^2$ (blue) and $|\psi_2|^2$ (red). The dynamic structure factors are frequency broadened by a frequency width of $\omega_B=0.1\omega_z$.
    	}
	\label{Fig:DSFunbalanced}
\end{figure}In Fig.~\ref{Fig:DSFunbalanced} we present results for the dynamic and static structure factors for two unbalanced cases. In the first case [Figs.~\ref{Fig:DSFunbalanced}(a1)-(a3)] the DDIs for the second component are half the strength of the first (i.e.~$g^{dd}_{22}=g^{dd}_{11}/2$).  In the second case [Figs.~\ref{Fig:DSFunbalanced}(b1)-(b3)] the second component is non-dipolar  (i.e.~$g^{dd}_{22}=0$).  These results can be considered as a progression from the results of Figs.~\ref{Fig:DSF}(a1)-(a3), in which the DDIs of the second component are reduced, but the other parameters are adjusted to be close to the roton boundary. As these cases are unbalanced, the theory of Sec.~\ref{Sec:BalancedTheory} is inapplicable and the excitations do not separate into in-phase/density and out-of-phase/spin classes, so the theory of Secs.~\ref{s:mftheory} and \ref{s:excitations} is required.

The insets to Figs.~\ref{Fig:DSFunbalanced}(a3) and (b3) show that the axial condensate density profiles for these cases differ, as $g_{11}>g_{22}$, so the variational theory, which assumes that both components are the same, is no longer applicable. 

The rotons observed are now of mixed density and spin character, and they contribute weight to both the density and spin-density dynamic structure factors.  In practice, we identify the character of the roton by the dominant peak of the static structure factors. For example, the case in Fig.~\ref{Fig:DSFunbalanced}(a3) is a density-dominant roton (i.e.,~$S_+$ has a higher peak than $S_-$), whereas the case in Fig.~\ref{Fig:DSFunbalanced}(b3) is a spin-dominant roton.

\subsection{Stability phase diagram}

 \begin{figure}[htp!] 
	\centering
	\includegraphics[width=\linewidth]{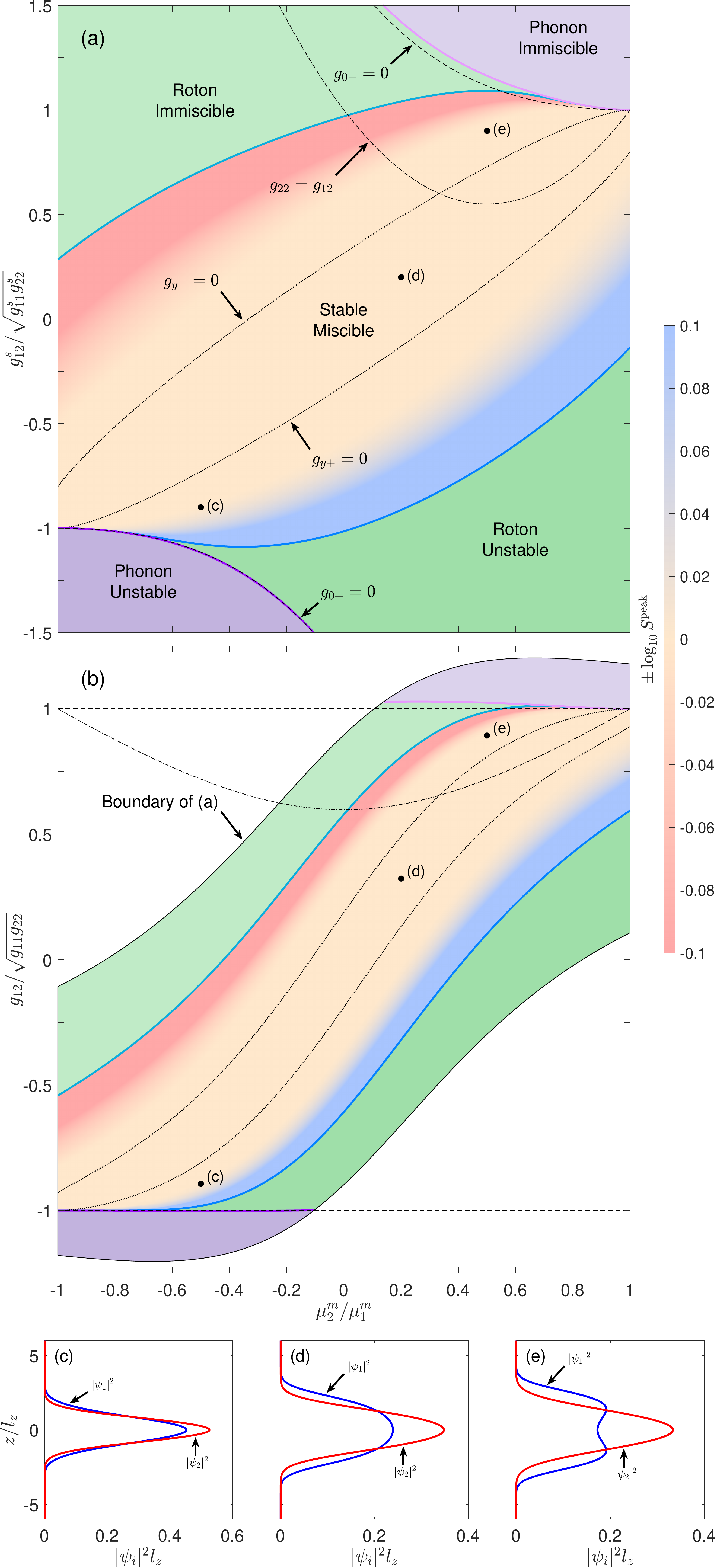}
    \caption{Stability phase diagrams for an unbalanced system with axes $\mu_2^m/\mu_1^m$, (a) $g^s_{12}/\sqrt{g^s_{11}g^s_{22}}$, and (b)  $g_{12}/\sqrt{g_{11}g_{22}}$. Color of the stable miscible region reveals the dominant character of the roton when present (see text). Blue (purple) boundaries identify when a roton (phonon) mode goes soft leading to an instability.   (c)-(e) Wavefunctions of the components at various points in the stable region as indicated in the main plot.
	Parameters: $ng_{ii}^s/\hbar\omega_zl_z= 5$, $ng_{11}^{dd}/\hbar\omega_zl_z= 4.5$ and $\alpha=0$.  }
	\label{Fig:unbalancedPD}
\end{figure}

A stability phase diagram for a general unbalanced case is shown in Fig.~\ref{Fig:unbalancedPD}(a). We have used $g^s_{12}/\sqrt{g_{11}^sg_{22}^s}$ as the vertical axis of the phase diagram, which characterizes the stability for a two-component non-dipolar condensate\footnote{\label{f:homog}A homogeneous binary non-dipolar condensate is unstable to collapse for $g^s_{12}<-\sqrt{g_{11}^sg_{22}^s}$, while for $g^s_{12}>\sqrt{g_{11}^sg_{22}^s}$ it is immiscible.}. Along the horizontal axis we vary the ratio of the dipole components, with the left and right limits being the anti-parallel and parallel cases.

Similar to the procedure for the balanced case, we identify the instabilities from numerical calculations of the excitation spectrum.  A significant difference, as discussed in the last subsection, is that for the unbalanced case we need to use the structure factors to identify whether any dynamically unstable modes have dominant density (i.e.~unstable regions) or spin character (i.e.~immiscible regions).  
The boundaries are found using a square tracing algorithm \cite{Kovalevskii2021a}, and we refine results using bisection.  We note that the ground state in the stable miscible regime is always symmetric along $z$ for the parameters we consider in this phase diagram  (cf~to Fig.~2 of Ref.~\cite{Lee2021b}, also see Ref.~\cite{Gordon1998a,Chui1999a}).

In the stable region blue and red shading is used to quantify the dominant roton peak in the $S_+$ or $S_-$ structure factor, represented by $\log_{10}S^\mathrm{peak}\equiv\log_{10}\max_{\bk_\rho,\pm}S_\pm(\bk_\rho)$. We apply a negative sign if the peak of $S_-$ is greater than the peak of $S_+$, and if there is no peak we set the overall result to zero. Thus the upper red region indicates a spin-dominant roton, whereas the lower blue region indicates a density-dominant roton. These regions occur close to the boundaries to the portions of the phase diagram that are roton immiscible and roton unstable, respectively.

The phonon unstable region occurs in the lower left part of the phase diagram for an anti-parallel mixture. The phonon immiscible region occurs in the upper right part for a parallel-mixture. The criteria for these phonon regions is similar to the non-dipolar case\footnote{see footnote~\ref{f:homog}}, demonstrating that systems with parallel dipoles, the transition to immiscibility is largely unaffected by dipolar physics, while for anti-parallel dipoles the mechanical stability is largely unaffected (also see \cite{Lee2021b}).
 
 In Fig.~\ref{Fig:unbalancedPD}(b) we show the same stability phase diagram using $g_{12}/\sqrt{g_{11}g_{22}}$ as the vertical axis, emphasising the $g_{0\lambda}=0$ approximations to the immiscible (upper) and unstable (lower) phonon boundaries.

\subsection{Approximate stability criteria \label{s:approxstabcrit}} 
For the balanced system, the interaction potential $\Ut_\lambda$ introduced in Eq.~(\ref{Ulambda}) appears in the BdG equations \eqref{Xf}, and the $k_\rho\to 0$ and $k_y\to\infty$ limits of $\Ut_\lambda$ characterize the phonon instability and immiscibility boundaries, and provide lower bounds for the onset of the roton boundaries. In the unbalanced case, the lack of symmetry between the two components, $\psi_1\ne\psi_2$, means that there is no associated decoupling into in-phase and out-of-phase modes. Nevertheless we have found empirically that the $\Ut_\lambda$-potential given by Eq.~\eqref{Ulambda} is useful for the unbalanced case, where we still interpret $\lambda=1$ and $-1$ as indicating the effective density and spin interactions, respectively. Following the analysis of the balanced case we can find values for the effective low momentum and high momentum interactions as
${g}_{0\lambda}=\lim_{k_\rho\to0}\Ut_\lambda(\mathbf{k})$  and ${g}_{y\lambda}=\lim_{k_y\to\infty}\Ut_\lambda(\mathbf{k})$, and these quantities are given in Table \ref{tab:Utilde}.
Results using these limits for estimating the boundaries are shown in Figs.~\ref{Fig:unbalancedPD}(a) and (b).
Generally we have found that the analytical results provide close lower bounds when $\psi_1\approx\psi_2$. 
For example, $g_{0+}=0$ accurately predicts the phonon instability boundary, as in this region $\psi_1\approx\psi_2$ [Fig.~\ref{Fig:unbalancedPD}(c)]. 
Conversely, although $g_{0-}$ is qualitatively correct, it is less accurate since $\psi_1\not\approx\psi_2$. In particular there is a dip in the component with the highest interaction coupling constant, $\psi_1$ [see Fig.~\ref{Fig:unbalancedPD}(e)]\footnote{The appearance of a dip in the density of one component leads to the immiscibility curve given by the solid orange lines in Fig.~2 of \cite{Lee2021b}.}. For a non-dipolar system, within the Thomas-Fermi approximation, this dip would appear when $g_{22} < g_{12}$ [see \cite{Trippenbach2000a}, above the $g_{22}=g_{12}$ curve in Fig.~\ref{Fig:unbalancedPD}(a)]\footnote{Similarly when $g_{11}<g_{12}$, $|\psi_2|^2$ has a dip. In Fig.~\ref{Fig:unbalancedPD}, $g_{22}^{dd}\le g_{11}^{dd}$, so the $g_{11}=g_{12}$ curve would be within the phonon immiscible region for our parameters, and can be ignored.}.
 
As the analytical results are based on interactions only, the agreement between numerical and analytical results is best when interactions are dominant, i.e. $ng_{ij}\gg\hbar\omega_z$\footnote{For our choice of axes in Fig.~\ref{Fig:unbalancedPD}(a), the analytical boundaries remain unchanged if we change $g_{ii}^s$ and $g_{11}^{dd}$, keeping $g_{11}^s=g_{22}^s$, $\alpha=0$, and fix the ratio $g_{ii}^s/g_{11}^{dd}$.}.

We have compared our planar results to those for a system in a three-dimensional harmonic trap. 
We have chosen  $ng_{ii}^s/\hbar\omega_zl_z=5$ in Fig.~\ref{Fig:unbalancedPD} to approximately match the value of this quantity in Fig.~2(c) of Ref.~\cite{Lee2021b} at $\mu_2^m=-\mu_1^m$, and $g_{12}^s/g_{ii}^s=40/140$ (using the peak areal density from the trapped system for $n$).
There is good qualitative agreement between these two figures for the predictions of the stable region and nature of the instability, noting that quantitative agreement is not expected as the areal density of the trapped system varies over the phase diagram and across the harmonic confinement.

\section{Conclusions and outlook}\label{Sec:conclusion}
In this paper we have developed theory for a two-component miscible dipolar condensate in a planar trap. Our theory allows us to quantify the various instabilities arising because of phonon or roton modes becoming dynamically unstable. Notably we  develop a number of analytic results that characterize the boundaries  or provide lower bounds for these instabilities to develop. The roton unstable and roton immiscible regimes emerge as the new features from the DDIs under confinement. We have also explored how the roton excitations driving these instabilities are revealed through the density- or spin-density fluctuations, as characterized by the dynamic and static structure factors.

  In this work we do not explore the nature of the new ground states that would emerge from these instabilities. Where the instability is of a density type (i.e.~phonon or roton unstable) the system is expected to mechanically collapse or implode. In this case quantum fluctuation effects can stabilize the collapse leading to a stable higher density quantum droplet array or a density supersolid state forming. Where the instability is of the spin-density type (i.e.~phonon or roton immiscible), we instead expect phase separation of the two components to occur. For the case of an unstable spin-roton the ground state in this regime may be an immiscible pattern characterized by a finite microscopic length-scale (e.g., striped pattern or two-dimensional crystal). Aspects of roton-immiscibility dynamics and ground states have been  explored in a dipolar-non-dipolar mixture in planar systems \cite{Saito2009,Wilson2012a}. This is also a regime where a domain-supersolid can develop in which the two components form a series of alternating domains, producing an immiscible double supersolid (e.g.~see \cite{Li2022a,Bland2022a}). Interestingly, this can occur at lower densities where the quantum fluctuation effects are not significant and loss rates will be much lower.   Our results show that the anti-parallel system might   be  favourable for studying the roton-immiscibility transition and that it can occur in a broad ranges of cases for a suitable choice of interaction parameters.

Here we have focused on the planar system, however our approach can be extended to the infinite tubular regime that provides a model of the cigar shape traps that have been extensively used in recent dipolar Bose-Einstein condensate experiments exploring rotons and supersolids (e.g.~see \cite{Chomaz2018a,Tanzi2018,Bottcher2019,Chomaz2019}).   
 
\subsection* {Acknowledgments}
Useful discussions with R.~Bisset and support from the Marsden Fund of the Royal Society of New Zealand are acknowledged.


\begin{thebibliography}{37}%
\makeatletter
\providecommand \@ifxundefined [1]{%
 \@ifx{#1\undefined}
}%
\providecommand \@ifnum [1]{%
 \ifnum #1\expandafter \@firstoftwo
 \else \expandafter \@secondoftwo
 \fi
}%
\providecommand \@ifx [1]{%
 \ifx #1\expandafter \@firstoftwo
 \else \expandafter \@secondoftwo
 \fi
}%
\providecommand \natexlab [1]{#1}%
\providecommand \enquote  [1]{``#1''}%
\providecommand \bibnamefont  [1]{#1}%
\providecommand \bibfnamefont [1]{#1}%
\providecommand \citenamefont [1]{#1}%
\providecommand \href@noop [0]{\@secondoftwo}%
\providecommand \href [0]{\begingroup \@sanitize@url \@href}%
\providecommand \@href[1]{\@@startlink{#1}\@@href}%
\providecommand \@@href[1]{\endgroup#1\@@endlink}%
\providecommand \@sanitize@url [0]{\catcode `\\12\catcode `\$12\catcode
  `\&12\catcode `\#12\catcode `\^12\catcode `\_12\catcode `\%12\relax}%
\providecommand \@@startlink[1]{}%
\providecommand \@@endlink[0]{}%
\providecommand \url  [0]{\begingroup\@sanitize@url \@url }%
\providecommand \@url [1]{\endgroup\@href {#1}{\urlprefix }}%
\providecommand \urlprefix  [0]{URL }%
\providecommand \Eprint [0]{\href }%
\providecommand \doibase [0]{https://doi.org/}%
\providecommand \selectlanguage [0]{\@gobble}%
\providecommand \bibinfo  [0]{\@secondoftwo}%
\providecommand \bibfield  [0]{\@secondoftwo}%
\providecommand \translation [1]{[#1]}%
\providecommand \BibitemOpen [0]{}%
\providecommand \bibitemStop [0]{}%
\providecommand \bibitemNoStop [0]{.\EOS\space}%
\providecommand \EOS [0]{\spacefactor3000\relax}%
\providecommand \BibitemShut  [1]{\csname bibitem#1\endcsname}%
\let\auto@bib@innerbib\@empty
\bibitem [{\citenamefont {Trautmann}\ \emph {et~al.}(2018)\citenamefont
  {Trautmann}, \citenamefont {Ilzh\"ofer}, \citenamefont {Durastante},
  \citenamefont {Politi}, \citenamefont {Sohmen}, \citenamefont {Mark},\ and\
  \citenamefont {Ferlaino}}]{Trautmann2018}%
  \BibitemOpen
  \bibfield  {author} {\bibinfo {author} {\bibfnamefont {A.}~\bibnamefont
  {Trautmann}}, \bibinfo {author} {\bibfnamefont {P.}~\bibnamefont
  {Ilzh\"ofer}}, \bibinfo {author} {\bibfnamefont {G.}~\bibnamefont
  {Durastante}}, \bibinfo {author} {\bibfnamefont {C.}~\bibnamefont {Politi}},
  \bibinfo {author} {\bibfnamefont {M.}~\bibnamefont {Sohmen}}, \bibinfo
  {author} {\bibfnamefont {M.~J.}\ \bibnamefont {Mark}},\ and\ \bibinfo
  {author} {\bibfnamefont {F.}~\bibnamefont {Ferlaino}},\ }\bibfield  {title}
  {\bibinfo {title} {Dipolar quantum mixtures of erbium and dysprosium atoms},\
  }\href {https://doi.org/10.1103/PhysRevLett.121.213601} {\bibfield  {journal}
  {\bibinfo  {journal} {Phys. Rev. Lett.}\ }\textbf {\bibinfo {volume} {121}},\
  \bibinfo {pages} {213601} (\bibinfo {year} {2018})}\BibitemShut {NoStop}%
\bibitem [{\citenamefont {Politi}\ \emph {et~al.}(2022)\citenamefont {Politi},
  \citenamefont {Trautmann}, \citenamefont {Ilzh\"ofer}, \citenamefont
  {Durastante}, \citenamefont {Mark}, \citenamefont {Modugno},\ and\
  \citenamefont {Ferlaino}}]{Politi2022a}%
  \BibitemOpen
  \bibfield  {author} {\bibinfo {author} {\bibfnamefont {C.}~\bibnamefont
  {Politi}}, \bibinfo {author} {\bibfnamefont {A.}~\bibnamefont {Trautmann}},
  \bibinfo {author} {\bibfnamefont {P.}~\bibnamefont {Ilzh\"ofer}}, \bibinfo
  {author} {\bibfnamefont {G.}~\bibnamefont {Durastante}}, \bibinfo {author}
  {\bibfnamefont {M.~J.}\ \bibnamefont {Mark}}, \bibinfo {author}
  {\bibfnamefont {M.}~\bibnamefont {Modugno}},\ and\ \bibinfo {author}
  {\bibfnamefont {F.}~\bibnamefont {Ferlaino}},\ }\bibfield  {title} {\bibinfo
  {title} {Interspecies interactions in an ultracold dipolar mixture},\ }\href
  {https://doi.org/10.1103/PhysRevA.105.023304} {\bibfield  {journal} {\bibinfo
   {journal} {Phys. Rev. A}\ }\textbf {\bibinfo {volume} {105}},\ \bibinfo
  {pages} {023304} (\bibinfo {year} {2022})}\BibitemShut {NoStop}%
\bibitem [{\citenamefont {Durastante}\ \emph {et~al.}(2020)\citenamefont
  {Durastante}, \citenamefont {Politi}, \citenamefont {Sohmen}, \citenamefont
  {Ilzh\"ofer}, \citenamefont {Mark}, \citenamefont {Norcia},\ and\
  \citenamefont {Ferlaino}}]{Durastante2020}%
  \BibitemOpen
  \bibfield  {author} {\bibinfo {author} {\bibfnamefont {G.}~\bibnamefont
  {Durastante}}, \bibinfo {author} {\bibfnamefont {C.}~\bibnamefont {Politi}},
  \bibinfo {author} {\bibfnamefont {M.}~\bibnamefont {Sohmen}}, \bibinfo
  {author} {\bibfnamefont {P.}~\bibnamefont {Ilzh\"ofer}}, \bibinfo {author}
  {\bibfnamefont {M.~J.}\ \bibnamefont {Mark}}, \bibinfo {author}
  {\bibfnamefont {M.~A.}\ \bibnamefont {Norcia}},\ and\ \bibinfo {author}
  {\bibfnamefont {F.}~\bibnamefont {Ferlaino}},\ }\bibfield  {title} {\bibinfo
  {title} {Feshbach resonances in an erbium-dysprosium dipolar mixture},\
  }\href {https://doi.org/10.1103/PhysRevA.102.033330} {\bibfield  {journal}
  {\bibinfo  {journal} {Phys. Rev. A}\ }\textbf {\bibinfo {volume} {102}},\
  \bibinfo {pages} {033330} (\bibinfo {year} {2020})}\BibitemShut {NoStop}%
\bibitem [{\citenamefont {Chalopin}\ \emph {et~al.}(2020)\citenamefont
  {Chalopin}, \citenamefont {Satoor}, \citenamefont {Evrard}, \citenamefont
  {Makhalov}, \citenamefont {Dalibard}, \citenamefont {Lopes},\ and\
  \citenamefont {Nascimbene}}]{Chalopin2020}%
  \BibitemOpen
  \bibfield  {author} {\bibinfo {author} {\bibfnamefont {T.}~\bibnamefont
  {Chalopin}}, \bibinfo {author} {\bibfnamefont {T.}~\bibnamefont {Satoor}},
  \bibinfo {author} {\bibfnamefont {A.}~\bibnamefont {Evrard}}, \bibinfo
  {author} {\bibfnamefont {V.}~\bibnamefont {Makhalov}}, \bibinfo {author}
  {\bibfnamefont {J.}~\bibnamefont {Dalibard}}, \bibinfo {author}
  {\bibfnamefont {R.}~\bibnamefont {Lopes}},\ and\ \bibinfo {author}
  {\bibfnamefont {S.}~\bibnamefont {Nascimbene}},\ }\bibfield  {title}
  {\bibinfo {title} {Probing chiral edge dynamics and bulk topology of a
  synthetic {H}all system},\ }\href {https://doi.org/10.1038/s41567-020-0942-5}
  {\bibfield  {journal} {\bibinfo  {journal} {Nat. Phys.}\ }\textbf {\bibinfo
  {volume} {16}},\ \bibinfo {pages} {1017} (\bibinfo {year}
  {2020})}\BibitemShut {NoStop}%
\bibitem [{\citenamefont {Ravensbergen}\ \emph {et~al.}(2018)\citenamefont
  {Ravensbergen}, \citenamefont {Corre}, \citenamefont {Soave}, \citenamefont
  {Kreyer}, \citenamefont {Kirilov},\ and\ \citenamefont
  {Grimm}}]{Ravensbergen2018a}%
  \BibitemOpen
  \bibfield  {author} {\bibinfo {author} {\bibfnamefont {C.}~\bibnamefont
  {Ravensbergen}}, \bibinfo {author} {\bibfnamefont {V.}~\bibnamefont {Corre}},
  \bibinfo {author} {\bibfnamefont {E.}~\bibnamefont {Soave}}, \bibinfo
  {author} {\bibfnamefont {M.}~\bibnamefont {Kreyer}}, \bibinfo {author}
  {\bibfnamefont {E.}~\bibnamefont {Kirilov}},\ and\ \bibinfo {author}
  {\bibfnamefont {R.}~\bibnamefont {Grimm}},\ }\bibfield  {title} {\bibinfo
  {title} {Production of a degenerate {F}ermi-{F}ermi mixture of dysprosium and
  potassium atoms},\ }\href {https://doi.org/10.1103/PhysRevA.98.063624}
  {\bibfield  {journal} {\bibinfo  {journal} {Phys. Rev. A}\ }\textbf {\bibinfo
  {volume} {98}},\ \bibinfo {pages} {063624} (\bibinfo {year}
  {2018})}\BibitemShut {NoStop}%
\bibitem [{\citenamefont {Ravensbergen}\ \emph {et~al.}(2020)\citenamefont
  {Ravensbergen}, \citenamefont {Soave}, \citenamefont {Corre}, \citenamefont
  {Kreyer}, \citenamefont {Huang}, \citenamefont {Kirilov},\ and\ \citenamefont
  {Grimm}}]{Ravensbergen2020a}%
  \BibitemOpen
  \bibfield  {author} {\bibinfo {author} {\bibfnamefont {C.}~\bibnamefont
  {Ravensbergen}}, \bibinfo {author} {\bibfnamefont {E.}~\bibnamefont {Soave}},
  \bibinfo {author} {\bibfnamefont {V.}~\bibnamefont {Corre}}, \bibinfo
  {author} {\bibfnamefont {M.}~\bibnamefont {Kreyer}}, \bibinfo {author}
  {\bibfnamefont {B.}~\bibnamefont {Huang}}, \bibinfo {author} {\bibfnamefont
  {E.}~\bibnamefont {Kirilov}},\ and\ \bibinfo {author} {\bibfnamefont
  {R.}~\bibnamefont {Grimm}},\ }\bibfield  {title} {\bibinfo {title}
  {Resonantly interacting {F}ermi-{F}ermi mixture of $^{161}\mathrm{Dy}$ and
  $^{40}\mathrm{K}$},\ }\href {https://doi.org/10.1103/PhysRevLett.124.203402}
  {\bibfield  {journal} {\bibinfo  {journal} {Phys. Rev. Lett.}\ }\textbf
  {\bibinfo {volume} {124}},\ \bibinfo {pages} {203402} (\bibinfo {year}
  {2020})}\BibitemShut {NoStop}%
\bibitem [{\citenamefont {Bisset}\ \emph {et~al.}(2021)\citenamefont {Bisset},
  \citenamefont {Ardila},\ and\ \citenamefont {Santos}}]{Bisset2021}%
  \BibitemOpen
  \bibfield  {author} {\bibinfo {author} {\bibfnamefont {R.~N.}\ \bibnamefont
  {Bisset}}, \bibinfo {author} {\bibfnamefont {L.~A.~P.}\ \bibnamefont
  {Ardila}},\ and\ \bibinfo {author} {\bibfnamefont {L.}~\bibnamefont
  {Santos}},\ }\bibfield  {title} {\bibinfo {title} {Quantum droplets of
  dipolar mixtures},\ }\href {https://doi.org/10.1103/PhysRevLett.126.025301}
  {\bibfield  {journal} {\bibinfo  {journal} {Phys. Rev. Lett.}\ }\textbf
  {\bibinfo {volume} {126}},\ \bibinfo {pages} {025301} (\bibinfo {year}
  {2021})}\BibitemShut {NoStop}%
\bibitem [{\citenamefont {Smith}\ \emph
  {et~al.}(2021{\natexlab{a}})\citenamefont {Smith}, \citenamefont {Baillie},\
  and\ \citenamefont {Blakie}}]{Smith2021a}%
  \BibitemOpen
  \bibfield  {author} {\bibinfo {author} {\bibfnamefont {J.~C.}\ \bibnamefont
  {Smith}}, \bibinfo {author} {\bibfnamefont {D.}~\bibnamefont {Baillie}},\
  and\ \bibinfo {author} {\bibfnamefont {P.~B.}\ \bibnamefont {Blakie}},\
  }\bibfield  {title} {\bibinfo {title} {Quantum droplet states of a binary
  magnetic gas},\ }\href {https://doi.org/10.1103/PhysRevLett.126.025302}
  {\bibfield  {journal} {\bibinfo  {journal} {Phys. Rev. Lett.}\ }\textbf
  {\bibinfo {volume} {126}},\ \bibinfo {pages} {025302} (\bibinfo {year}
  {2021}{\natexlab{a}})}\BibitemShut {NoStop}%
\bibitem [{\citenamefont {Smith}\ \emph
  {et~al.}(2021{\natexlab{b}})\citenamefont {Smith}, \citenamefont {Blakie},\
  and\ \citenamefont {Baillie}}]{Smith2021b}%
  \BibitemOpen
  \bibfield  {author} {\bibinfo {author} {\bibfnamefont {J.~C.}\ \bibnamefont
  {Smith}}, \bibinfo {author} {\bibfnamefont {P.~B.}\ \bibnamefont {Blakie}},\
  and\ \bibinfo {author} {\bibfnamefont {D.}~\bibnamefont {Baillie}},\
  }\bibfield  {title} {\bibinfo {title} {Approximate theories for binary
  magnetic quantum droplets},\ }\href
  {https://doi.org/10.1103/PhysRevA.104.053316} {\bibfield  {journal} {\bibinfo
   {journal} {Phys. Rev. A}\ }\textbf {\bibinfo {volume} {104}},\ \bibinfo
  {pages} {053316} (\bibinfo {year} {2021}{\natexlab{b}})}\BibitemShut
  {NoStop}%
\bibitem [{\citenamefont {Saito}\ \emph {et~al.}(2009)\citenamefont {Saito},
  \citenamefont {Kawaguchi},\ and\ \citenamefont {Ueda}}]{Saito2009}%
  \BibitemOpen
  \bibfield  {author} {\bibinfo {author} {\bibfnamefont {H.}~\bibnamefont
  {Saito}}, \bibinfo {author} {\bibfnamefont {Y.}~\bibnamefont {Kawaguchi}},\
  and\ \bibinfo {author} {\bibfnamefont {M.}~\bibnamefont {Ueda}},\ }\bibfield
  {title} {\bibinfo {title} {Ferrofluidity in a two-component dipolar
  {B}ose-{E}instein condensate},\ }\href
  {https://doi.org/10.1103/PhysRevLett.102.230403} {\bibfield  {journal}
  {\bibinfo  {journal} {Phys. Rev. Lett.}\ }\textbf {\bibinfo {volume} {102}},\
  \bibinfo {pages} {230403} (\bibinfo {year} {2009})}\BibitemShut {NoStop}%
\bibitem [{\citenamefont {Scheiermann}\ \emph {et~al.}()\citenamefont
  {Scheiermann}, \citenamefont {Ardila}, \citenamefont {Bland}, \citenamefont
  {Bisset},\ and\ \citenamefont {Santos}}]{Scheiermann2022a}%
  \BibitemOpen
  \bibfield  {author} {\bibinfo {author} {\bibfnamefont {D.}~\bibnamefont
  {Scheiermann}}, \bibinfo {author} {\bibfnamefont {L.~A.~P.}\ \bibnamefont
  {Ardila}}, \bibinfo {author} {\bibfnamefont {T.}~\bibnamefont {Bland}},
  \bibinfo {author} {\bibfnamefont {R.~N.}\ \bibnamefont {Bisset}},\ and\
  \bibinfo {author} {\bibfnamefont {L.}~\bibnamefont {Santos}},\ }\bibfield
  {title} {\bibinfo {title} {Catalyzation of supersolidity in binary dipolar
  condensates},\ }\href@noop {} {\ }\Eprint {https://arxiv.org/abs/2202.08259}
  {arXiv:2202.08259} \BibitemShut {NoStop}%
\bibitem [{\citenamefont {Li}\ \emph {et~al.}()\citenamefont {Li},
  \citenamefont {Le},\ and\ \citenamefont {Saito}}]{Li2022a}%
  \BibitemOpen
  \bibfield  {author} {\bibinfo {author} {\bibfnamefont {S.}~\bibnamefont
  {Li}}, \bibinfo {author} {\bibfnamefont {U.~N.}\ \bibnamefont {Le}},\ and\
  \bibinfo {author} {\bibfnamefont {H.}~\bibnamefont {Saito}},\ }\bibfield
  {title} {\bibinfo {title} {Long lifetime supersolid in a two-component
  dipolar {B}ose-{E}instein condensate},\ }\href@noop {} {\ }\Eprint
  {https://arxiv.org/abs/2203.09799} {arXiv:2203.09799} \BibitemShut {NoStop}%
\bibitem [{\citenamefont {Bland}\ \emph {et~al.}()\citenamefont {Bland},
  \citenamefont {Poli}, \citenamefont {Ardila}, \citenamefont {Santos},
  \citenamefont {Ferlaino},\ and\ \citenamefont {Bisset}}]{Bland2022a}%
  \BibitemOpen
  \bibfield  {author} {\bibinfo {author} {\bibfnamefont {T.}~\bibnamefont
  {Bland}}, \bibinfo {author} {\bibfnamefont {E.}~\bibnamefont {Poli}},
  \bibinfo {author} {\bibfnamefont {L.~A.~P.}\ \bibnamefont {Ardila}}, \bibinfo
  {author} {\bibfnamefont {L.}~\bibnamefont {Santos}}, \bibinfo {author}
  {\bibfnamefont {F.}~\bibnamefont {Ferlaino}},\ and\ \bibinfo {author}
  {\bibfnamefont {R.~N.}\ \bibnamefont {Bisset}},\ }\bibfield  {title}
  {\bibinfo {title} {Domain supersolids in binary dipolar condensates},\
  }\href@noop {} {\ }\Eprint {https://arxiv.org/abs/2203.11119}
  {arXiv:2203.11119} \BibitemShut {NoStop}%
\bibitem [{\citenamefont {Wilson}\ \emph {et~al.}(2012)\citenamefont {Wilson},
  \citenamefont {Ticknor}, \citenamefont {Bohn},\ and\ \citenamefont
  {Timmermans}}]{Wilson2012a}%
  \BibitemOpen
  \bibfield  {author} {\bibinfo {author} {\bibfnamefont {R.~M.}\ \bibnamefont
  {Wilson}}, \bibinfo {author} {\bibfnamefont {C.}~\bibnamefont {Ticknor}},
  \bibinfo {author} {\bibfnamefont {J.~L.}\ \bibnamefont {Bohn}},\ and\
  \bibinfo {author} {\bibfnamefont {E.}~\bibnamefont {Timmermans}},\ }\bibfield
   {title} {\bibinfo {title} {Roton immiscibility in a two-component dipolar
  {Bose} gas},\ }\href {https://doi.org/10.1103/PhysRevA.86.033606} {\bibfield
  {journal} {\bibinfo  {journal} {Phys. Rev. A}\ }\textbf {\bibinfo {volume}
  {86}},\ \bibinfo {pages} {033606} (\bibinfo {year} {2012})}\BibitemShut
  {NoStop}%
\bibitem [{\citenamefont {Xi}\ \emph {et~al.}(2018)\citenamefont {Xi},
  \citenamefont {Byrnes},\ and\ \citenamefont {Saito}}]{KuiTian2018}%
  \BibitemOpen
  \bibfield  {author} {\bibinfo {author} {\bibfnamefont {K.-T.}\ \bibnamefont
  {Xi}}, \bibinfo {author} {\bibfnamefont {T.}~\bibnamefont {Byrnes}},\ and\
  \bibinfo {author} {\bibfnamefont {H.}~\bibnamefont {Saito}},\ }\bibfield
  {title} {\bibinfo {title} {Fingering instabilities and pattern formation in a
  two-component dipolar {Bose}-{Einstein} condensate},\ }\href
  {https://doi.org/10.1103/PhysRevA.97.023625} {\bibfield  {journal} {\bibinfo
  {journal} {Phys. Rev. A}\ }\textbf {\bibinfo {volume} {97}},\ \bibinfo
  {pages} {023625} (\bibinfo {year} {2018})}\BibitemShut {NoStop}%
\bibitem [{\citenamefont {Lee}\ \emph {et~al.}(2021{\natexlab{a}})\citenamefont
  {Lee}, \citenamefont {Baillie}, \citenamefont {Blakie},\ and\ \citenamefont
  {Bisset}}]{Lee2021b}%
  \BibitemOpen
  \bibfield  {author} {\bibinfo {author} {\bibfnamefont {A.-C.}\ \bibnamefont
  {Lee}}, \bibinfo {author} {\bibfnamefont {D.}~\bibnamefont {Baillie}},
  \bibinfo {author} {\bibfnamefont {P.~B.}\ \bibnamefont {Blakie}},\ and\
  \bibinfo {author} {\bibfnamefont {R.~N.}\ \bibnamefont {Bisset}},\ }\bibfield
   {title} {\bibinfo {title} {Miscibility and stability of dipolar bosonic
  mixtures},\ }\href {https://doi.org/10.1103/PhysRevA.103.063301} {\bibfield
  {journal} {\bibinfo  {journal} {Phys. Rev. A}\ }\textbf {\bibinfo {volume}
  {103}},\ \bibinfo {pages} {063301} (\bibinfo {year}
  {2021}{\natexlab{a}})}\BibitemShut {NoStop}%
\bibitem [{\citenamefont {Zhang}\ \emph {et~al.}(2015)\citenamefont {Zhang},
  \citenamefont {Han}, \citenamefont {Wen}, \citenamefont {Zhang},
  \citenamefont {Dong}, \citenamefont {Chang},\ and\ \citenamefont
  {Zhang}}]{Zhang2015}%
  \BibitemOpen
  \bibfield  {author} {\bibinfo {author} {\bibfnamefont {X.-F.}\ \bibnamefont
  {Zhang}}, \bibinfo {author} {\bibfnamefont {W.}~\bibnamefont {Han}}, \bibinfo
  {author} {\bibfnamefont {L.}~\bibnamefont {Wen}}, \bibinfo {author}
  {\bibfnamefont {P.}~\bibnamefont {Zhang}}, \bibinfo {author} {\bibfnamefont
  {R.-F.}\ \bibnamefont {Dong}}, \bibinfo {author} {\bibfnamefont
  {H.}~\bibnamefont {Chang}},\ and\ \bibinfo {author} {\bibfnamefont {S.-G.}\
  \bibnamefont {Zhang}},\ }\bibfield  {title} {\bibinfo {title} {Two-component
  dipolar {B}ose-{E}instein condensate in concentrically coupled annular
  traps},\ }\href {https://doi.org/10.1038/srep08684} {\bibfield  {journal}
  {\bibinfo  {journal} {Scientific Reports}\ }\textbf {\bibinfo {volume} {5}},\
  \bibinfo {pages} {8684} (\bibinfo {year} {2015})}\BibitemShut {NoStop}%
\bibitem [{\citenamefont {Zhang}\ \emph {et~al.}(2016)\citenamefont {Zhang},
  \citenamefont {Wen}, \citenamefont {Dai}, \citenamefont {Dong}, \citenamefont
  {Jiang}, \citenamefont {Chang},\ and\ \citenamefont {Zhang}}]{Zhang2016}%
  \BibitemOpen
  \bibfield  {author} {\bibinfo {author} {\bibfnamefont {X.-F.}\ \bibnamefont
  {Zhang}}, \bibinfo {author} {\bibfnamefont {L.}~\bibnamefont {Wen}}, \bibinfo
  {author} {\bibfnamefont {C.-Q.}\ \bibnamefont {Dai}}, \bibinfo {author}
  {\bibfnamefont {R.-F.}\ \bibnamefont {Dong}}, \bibinfo {author}
  {\bibfnamefont {H.-F.}\ \bibnamefont {Jiang}}, \bibinfo {author}
  {\bibfnamefont {H.}~\bibnamefont {Chang}},\ and\ \bibinfo {author}
  {\bibfnamefont {S.-G.}\ \bibnamefont {Zhang}},\ }\bibfield  {title} {\bibinfo
  {title} {Exotic vortex lattices in a rotating binary dipolar
  {Bose}-{Einstein} condensate},\ }\href {https://doi.org/10.1038/srep19380}
  {\bibfield  {journal} {\bibinfo  {journal} {Scientific Reports}\ }\textbf
  {\bibinfo {volume} {6}},\ \bibinfo {pages} {19380} (\bibinfo {year}
  {2016})}\BibitemShut {NoStop}%
\bibitem [{\citenamefont {Kumar}\ \emph {et~al.}(2017)\citenamefont {Kumar},
  \citenamefont {Tomio}, \citenamefont {Malomed},\ and\ \citenamefont
  {Gammal}}]{Kumar2017}%
  \BibitemOpen
  \bibfield  {author} {\bibinfo {author} {\bibfnamefont {R.~K.}\ \bibnamefont
  {Kumar}}, \bibinfo {author} {\bibfnamefont {L.}~\bibnamefont {Tomio}},
  \bibinfo {author} {\bibfnamefont {B.~A.}\ \bibnamefont {Malomed}},\ and\
  \bibinfo {author} {\bibfnamefont {A.}~\bibnamefont {Gammal}},\ }\bibfield
  {title} {\bibinfo {title} {Vortex lattices in binary {B}ose-{E}instein
  condensates with dipole-dipole interactions},\ }\href
  {https://doi.org/10.1103/PhysRevA.96.063624} {\bibfield  {journal} {\bibinfo
  {journal} {Phys. Rev. A}\ }\textbf {\bibinfo {volume} {96}},\ \bibinfo
  {pages} {063624} (\bibinfo {year} {2017})}\BibitemShut {NoStop}%
\bibitem [{\citenamefont {Shirley}\ \emph {et~al.}(2014)\citenamefont
  {Shirley}, \citenamefont {Anderson}, \citenamefont {Clark},\ and\
  \citenamefont {Wilson}}]{Shirley2014}%
  \BibitemOpen
  \bibfield  {author} {\bibinfo {author} {\bibfnamefont {W.~E.}\ \bibnamefont
  {Shirley}}, \bibinfo {author} {\bibfnamefont {B.~M.}\ \bibnamefont
  {Anderson}}, \bibinfo {author} {\bibfnamefont {C.~W.}\ \bibnamefont
  {Clark}},\ and\ \bibinfo {author} {\bibfnamefont {R.~M.}\ \bibnamefont
  {Wilson}},\ }\bibfield  {title} {\bibinfo {title} {Half-quantum vortex
  molecules in a binary dipolar {B}ose gas},\ }\href
  {https://doi.org/10.1103/PhysRevLett.113.165301} {\bibfield  {journal}
  {\bibinfo  {journal} {Phys. Rev. Lett.}\ }\textbf {\bibinfo {volume} {113}},\
  \bibinfo {pages} {165301} (\bibinfo {year} {2014})}\BibitemShut {NoStop}%
\bibitem [{\citenamefont {Pradas}\ and\ \citenamefont
  {Boronat}(2022)}]{Pradas2022}%
  \BibitemOpen
  \bibfield  {author} {\bibinfo {author} {\bibfnamefont {S.}~\bibnamefont
  {Pradas}}\ and\ \bibinfo {author} {\bibfnamefont {J.}~\bibnamefont
  {Boronat}},\ }\bibfield  {title} {\bibinfo {title} {Mixtures of dipolar gases
  in two dimensions: A quantum {M}onte {C}arlo study},\ }\href
  {https://doi.org/10.3390/condmat7020032} {\bibfield  {journal} {\bibinfo
  {journal} {Condens. Matter}\ }\textbf {\bibinfo {volume} {7}},\ \bibinfo
  {pages} {32} (\bibinfo {year} {2022})}\BibitemShut {NoStop}%
\bibitem [{\citenamefont {G\'oral}\ and\ \citenamefont
  {Santos}(2002)}]{Goral2002}%
  \BibitemOpen
  \bibfield  {author} {\bibinfo {author} {\bibfnamefont {K.}~\bibnamefont
  {G\'oral}}\ and\ \bibinfo {author} {\bibfnamefont {L.}~\bibnamefont
  {Santos}},\ }\bibfield  {title} {\bibinfo {title} {Ground state and
  elementary excitations of single and binary {B}ose-{E}instein condensates of
  trapped dipolar gases},\ }\href {https://doi.org/10.1103/PhysRevA.66.023613}
  {\bibfield  {journal} {\bibinfo  {journal} {Phys. Rev. A}\ }\textbf {\bibinfo
  {volume} {66}},\ \bibinfo {pages} {023613} (\bibinfo {year}
  {2002})}\BibitemShut {NoStop}%
\bibitem [{\citenamefont {Baillie}\ and\ \citenamefont
  {Blakie}(2015)}]{Baillie2015a}%
  \BibitemOpen
  \bibfield  {author} {\bibinfo {author} {\bibfnamefont {D.}~\bibnamefont
  {Baillie}}\ and\ \bibinfo {author} {\bibfnamefont {P.~B.}\ \bibnamefont
  {Blakie}},\ }\bibfield  {title} {\bibinfo {title} {A general theory of
  flattened dipolar condensates},\ }\href
  {http://stacks.iop.org/1367-2630/17/i=3/a=033028} {\bibfield  {journal}
  {\bibinfo  {journal} {New Journal of Physics}\ }\textbf {\bibinfo {volume}
  {17}},\ \bibinfo {pages} {033028} (\bibinfo {year} {2015})}\BibitemShut
  {NoStop}%
\bibitem [{\citenamefont {Ronen}\ \emph {et~al.}(2006)\citenamefont {Ronen},
  \citenamefont {Bortolotti},\ and\ \citenamefont {Bohn}}]{Ronen2006a}%
  \BibitemOpen
  \bibfield  {author} {\bibinfo {author} {\bibfnamefont {S.}~\bibnamefont
  {Ronen}}, \bibinfo {author} {\bibfnamefont {D.~C.~E.}\ \bibnamefont
  {Bortolotti}},\ and\ \bibinfo {author} {\bibfnamefont {J.~L.}\ \bibnamefont
  {Bohn}},\ }\bibfield  {title} {\bibinfo {title} {{B}ogoliubov modes of a
  dipolar condensate in a cylindrical trap},\ }\href
  {https://doi.org/10.1103/PhysRevA.74.013623} {\bibfield  {journal} {\bibinfo
  {journal} {Phys. Rev. A}\ }\textbf {\bibinfo {volume} {74}},\ \bibinfo
  {pages} {013623} (\bibinfo {year} {2006})}\BibitemShut {NoStop}%
\bibitem [{\citenamefont {Lee}\ \emph {et~al.}(2021{\natexlab{b}})\citenamefont
  {Lee}, \citenamefont {Baillie},\ and\ \citenamefont {Blakie}}]{Lee2021a}%
  \BibitemOpen
  \bibfield  {author} {\bibinfo {author} {\bibfnamefont {A.-C.}\ \bibnamefont
  {Lee}}, \bibinfo {author} {\bibfnamefont {D.}~\bibnamefont {Baillie}},\ and\
  \bibinfo {author} {\bibfnamefont {P.~B.}\ \bibnamefont {Blakie}},\ }\bibfield
   {title} {\bibinfo {title} {Numerical calculation of dipolar-quantum-droplet
  stationary states},\ }\href
  {https://doi.org/10.1103/PhysRevResearch.3.013283} {\bibfield  {journal}
  {\bibinfo  {journal} {Phys. Rev. Research}\ }\textbf {\bibinfo {volume}
  {3}},\ \bibinfo {pages} {013283} (\bibinfo {year}
  {2021}{\natexlab{b}})}\BibitemShut {NoStop}%
\bibitem [{\citenamefont {Petter}\ \emph {et~al.}(2019)\citenamefont {Petter},
  \citenamefont {Natale}, \citenamefont {van Bijnen}, \citenamefont
  {Patscheider}, \citenamefont {Mark}, \citenamefont {Chomaz},\ and\
  \citenamefont {Ferlaino}}]{Petter2019}%
  \BibitemOpen
  \bibfield  {author} {\bibinfo {author} {\bibfnamefont {D.}~\bibnamefont
  {Petter}}, \bibinfo {author} {\bibfnamefont {G.}~\bibnamefont {Natale}},
  \bibinfo {author} {\bibfnamefont {R.~M.~W.}\ \bibnamefont {van Bijnen}},
  \bibinfo {author} {\bibfnamefont {A.}~\bibnamefont {Patscheider}}, \bibinfo
  {author} {\bibfnamefont {M.~J.}\ \bibnamefont {Mark}}, \bibinfo {author}
  {\bibfnamefont {L.}~\bibnamefont {Chomaz}},\ and\ \bibinfo {author}
  {\bibfnamefont {F.}~\bibnamefont {Ferlaino}},\ }\bibfield  {title} {\bibinfo
  {title} {Probing the roton excitation spectrum of a stable dipolar {B}ose
  gas},\ }\href {https://doi.org/10.1103/PhysRevLett.122.183401} {\bibfield
  {journal} {\bibinfo  {journal} {Phys. Rev. Lett.}\ }\textbf {\bibinfo
  {volume} {122}},\ \bibinfo {pages} {183401} (\bibinfo {year}
  {2019})}\BibitemShut {NoStop}%
\bibitem [{\citenamefont {Petter}\ \emph {et~al.}(2021)\citenamefont {Petter},
  \citenamefont {Patscheider}, \citenamefont {Natale}, \citenamefont {Mark},
  \citenamefont {Baranov}, \citenamefont {van Bijnen}, \citenamefont
  {Roccuzzo}, \citenamefont {Recati}, \citenamefont {Blakie}, \citenamefont
  {Baillie}, \citenamefont {Chomaz},\ and\ \citenamefont
  {Ferlaino}}]{Petter2021a}%
  \BibitemOpen
  \bibfield  {author} {\bibinfo {author} {\bibfnamefont {D.}~\bibnamefont
  {Petter}}, \bibinfo {author} {\bibfnamefont {A.}~\bibnamefont {Patscheider}},
  \bibinfo {author} {\bibfnamefont {G.}~\bibnamefont {Natale}}, \bibinfo
  {author} {\bibfnamefont {M.~J.}\ \bibnamefont {Mark}}, \bibinfo {author}
  {\bibfnamefont {M.~A.}\ \bibnamefont {Baranov}}, \bibinfo {author}
  {\bibfnamefont {R.}~\bibnamefont {van Bijnen}}, \bibinfo {author}
  {\bibfnamefont {S.~M.}\ \bibnamefont {Roccuzzo}}, \bibinfo {author}
  {\bibfnamefont {A.}~\bibnamefont {Recati}}, \bibinfo {author} {\bibfnamefont
  {B.}~\bibnamefont {Blakie}}, \bibinfo {author} {\bibfnamefont
  {D.}~\bibnamefont {Baillie}}, \bibinfo {author} {\bibfnamefont
  {L.}~\bibnamefont {Chomaz}},\ and\ \bibinfo {author} {\bibfnamefont
  {F.}~\bibnamefont {Ferlaino}},\ }\bibfield  {title} {\bibinfo {title} {Bragg
  scattering of an ultracold dipolar gas across the phase transition from
  {B}ose-{E}instein condensate to supersolid in the free-particle regime},\
  }\href {https://doi.org/10.1103/PhysRevA.104.L011302} {\bibfield  {journal}
  {\bibinfo  {journal} {Phys. Rev. A}\ }\textbf {\bibinfo {volume} {104}},\
  \bibinfo {pages} {L011302} (\bibinfo {year} {2021})}\BibitemShut {NoStop}%
\bibitem [{\citenamefont {Blakie}\ \emph {et~al.}(2020)\citenamefont {Blakie},
  \citenamefont {Baillie},\ and\ \citenamefont {Pal}}]{Blakie2020a}%
  \BibitemOpen
  \bibfield  {author} {\bibinfo {author} {\bibfnamefont {P.~B.}\ \bibnamefont
  {Blakie}}, \bibinfo {author} {\bibfnamefont {D.}~\bibnamefont {Baillie}},\
  and\ \bibinfo {author} {\bibfnamefont {S.}~\bibnamefont {Pal}},\ }\bibfield
  {title} {\bibinfo {title} {Variational theory for the ground state and
  collective excitations of an elongated dipolar condensate},\ }\href
  {https://doi.org/10.1088/1572-9494/ab95fa} {\bibfield  {journal} {\bibinfo
  {journal} {Commun. Theor. Phys}\ }\textbf {\bibinfo {volume} {72}},\ \bibinfo
  {pages} {085501} (\bibinfo {year} {2020})}\BibitemShut {NoStop}%
\bibitem [{\citenamefont {Pal}\ \emph {et~al.}(2020)\citenamefont {Pal},
  \citenamefont {Baillie},\ and\ \citenamefont {Blakie}}]{Pal2020a}%
  \BibitemOpen
  \bibfield  {author} {\bibinfo {author} {\bibfnamefont {S.}~\bibnamefont
  {Pal}}, \bibinfo {author} {\bibfnamefont {D.}~\bibnamefont {Baillie}},\ and\
  \bibinfo {author} {\bibfnamefont {P.~B.}\ \bibnamefont {Blakie}},\ }\bibfield
   {title} {\bibinfo {title} {Excitations and number fluctuations in an
  elongated dipolar {B}ose-{E}instein condensate},\ }\href
  {https://doi.org/10.1103/PhysRevA.102.043306} {\bibfield  {journal} {\bibinfo
   {journal} {Phys. Rev. A}\ }\textbf {\bibinfo {volume} {102}},\ \bibinfo
  {pages} {043306} (\bibinfo {year} {2020})}\BibitemShut {NoStop}%
\bibitem [{\citenamefont {Kovalevski{\v{i}}}(2021)}]{Kovalevskii2021a}%
  \BibitemOpen
  \bibfield  {author} {\bibinfo {author} {\bibfnamefont {V.~A.}\ \bibnamefont
  {Kovalevski{\v{i}}}},\ }\href@noop {} {\emph {\bibinfo {title} {Image
  processing with cellular topology}}}\ (\bibinfo  {publisher} {Springer},\
  \bibinfo {address} {Singapore},\ \bibinfo {year} {2021})\BibitemShut
  {NoStop}%
\bibitem [{\citenamefont {Gordon}\ and\ \citenamefont
  {Savage}(1998)}]{Gordon1998a}%
  \BibitemOpen
  \bibfield  {author} {\bibinfo {author} {\bibfnamefont {D.}~\bibnamefont
  {Gordon}}\ and\ \bibinfo {author} {\bibfnamefont {C.~M.}\ \bibnamefont
  {Savage}},\ }\bibfield  {title} {\bibinfo {title} {Excitation spectrum and
  instability of a two-species {B}ose-{E}instein condensate},\ }\href
  {https://doi.org/10.1103/PhysRevA.58.1440} {\bibfield  {journal} {\bibinfo
  {journal} {Phys. Rev. A}\ }\textbf {\bibinfo {volume} {58}},\ \bibinfo
  {pages} {1440} (\bibinfo {year} {1998})}\BibitemShut {NoStop}%
\bibitem [{\citenamefont {Chui}\ and\ \citenamefont {Ao}(1999)}]{Chui1999a}%
  \BibitemOpen
  \bibfield  {author} {\bibinfo {author} {\bibfnamefont {S.~T.}\ \bibnamefont
  {Chui}}\ and\ \bibinfo {author} {\bibfnamefont {P.}~\bibnamefont {Ao}},\
  }\bibfield  {title} {\bibinfo {title} {Broken cylindrical symmetry in binary
  mixtures of {B}ose-{E}instein condensates},\ }\href
  {https://doi.org/10.1103/PhysRevA.59.1473} {\bibfield  {journal} {\bibinfo
  {journal} {Phys. Rev. A}\ }\textbf {\bibinfo {volume} {59}},\ \bibinfo
  {pages} {1473} (\bibinfo {year} {1999})}\BibitemShut {NoStop}%
\bibitem [{\citenamefont {Trippenbach}\ \emph {et~al.}(2000)\citenamefont
  {Trippenbach}, \citenamefont {G{\'{o}}ral}, \citenamefont {Rzazewski},
  \citenamefont {Malomed},\ and\ \citenamefont {Band}}]{Trippenbach2000a}%
  \BibitemOpen
  \bibfield  {author} {\bibinfo {author} {\bibfnamefont {M.}~\bibnamefont
  {Trippenbach}}, \bibinfo {author} {\bibfnamefont {K.}~\bibnamefont
  {G{\'{o}}ral}}, \bibinfo {author} {\bibfnamefont {K.}~\bibnamefont
  {Rzazewski}}, \bibinfo {author} {\bibfnamefont {B.}~\bibnamefont {Malomed}},\
  and\ \bibinfo {author} {\bibfnamefont {Y.~B.}\ \bibnamefont {Band}},\
  }\bibfield  {title} {\bibinfo {title} {Structure of binary {B}ose-{E}instein
  condensates},\ }\href {https://doi.org/10.1088/0953-4075/33/19/314}
  {\bibfield  {journal} {\bibinfo  {journal} {J. Phys. B: At. Mol. Opt. Phys.}\
  }\textbf {\bibinfo {volume} {33}},\ \bibinfo {pages} {4017} (\bibinfo {year}
  {2000})}\BibitemShut {NoStop}%
\bibitem [{\citenamefont {Chomaz}\ \emph {et~al.}(2018)\citenamefont {Chomaz},
  \citenamefont {van Bijnen}, \citenamefont {Petter}, \citenamefont {Faraoni},
  \citenamefont {Baier}, \citenamefont {Becher}, \citenamefont {Mark},
  \citenamefont {W{\"a}chtler}, \citenamefont {Santos},\ and\ \citenamefont
  {Ferlaino}}]{Chomaz2018a}%
  \BibitemOpen
  \bibfield  {author} {\bibinfo {author} {\bibfnamefont {L.}~\bibnamefont
  {Chomaz}}, \bibinfo {author} {\bibfnamefont {R.~M.~W.}\ \bibnamefont {van
  Bijnen}}, \bibinfo {author} {\bibfnamefont {D.}~\bibnamefont {Petter}},
  \bibinfo {author} {\bibfnamefont {G.}~\bibnamefont {Faraoni}}, \bibinfo
  {author} {\bibfnamefont {S.}~\bibnamefont {Baier}}, \bibinfo {author}
  {\bibfnamefont {J.~H.}\ \bibnamefont {Becher}}, \bibinfo {author}
  {\bibfnamefont {M.~J.}\ \bibnamefont {Mark}}, \bibinfo {author}
  {\bibfnamefont {F.}~\bibnamefont {W{\"a}chtler}}, \bibinfo {author}
  {\bibfnamefont {L.}~\bibnamefont {Santos}},\ and\ \bibinfo {author}
  {\bibfnamefont {F.}~\bibnamefont {Ferlaino}},\ }\bibfield  {title} {\bibinfo
  {title} {Observation of roton mode population in a dipolar quantum gas},\
  }\href {https://doi.org/10.1038/s41567-018-0054-7} {\bibfield  {journal}
  {\bibinfo  {journal} {Nature Physics}\ }\textbf {\bibinfo {volume} {14}},\
  \bibinfo {pages} {442} (\bibinfo {year} {2018})}\BibitemShut {NoStop}%
\bibitem [{\citenamefont {Tanzi}\ \emph {et~al.}(2019)\citenamefont {Tanzi},
  \citenamefont {Lucioni}, \citenamefont {Fam\`a}, \citenamefont {Catani},
  \citenamefont {Fioretti}, \citenamefont {Gabbanini}, \citenamefont {Bisset},
  \citenamefont {Santos},\ and\ \citenamefont {Modugno}}]{Tanzi2018}%
  \BibitemOpen
  \bibfield  {author} {\bibinfo {author} {\bibfnamefont {L.}~\bibnamefont
  {Tanzi}}, \bibinfo {author} {\bibfnamefont {E.}~\bibnamefont {Lucioni}},
  \bibinfo {author} {\bibfnamefont {F.}~\bibnamefont {Fam\`a}}, \bibinfo
  {author} {\bibfnamefont {J.}~\bibnamefont {Catani}}, \bibinfo {author}
  {\bibfnamefont {A.}~\bibnamefont {Fioretti}}, \bibinfo {author}
  {\bibfnamefont {C.}~\bibnamefont {Gabbanini}}, \bibinfo {author}
  {\bibfnamefont {R.~N.}\ \bibnamefont {Bisset}}, \bibinfo {author}
  {\bibfnamefont {L.}~\bibnamefont {Santos}},\ and\ \bibinfo {author}
  {\bibfnamefont {G.}~\bibnamefont {Modugno}},\ }\bibfield  {title} {\bibinfo
  {title} {Observation of a dipolar quantum gas with metastable supersolid
  properties},\ }\href {https://doi.org/10.1103/PhysRevLett.122.130405}
  {\bibfield  {journal} {\bibinfo  {journal} {Phys. Rev. Lett.}\ }\textbf
  {\bibinfo {volume} {122}},\ \bibinfo {pages} {130405} (\bibinfo {year}
  {2019})}\BibitemShut {NoStop}%
\bibitem [{\citenamefont {B\"ottcher}\ \emph {et~al.}(2019)\citenamefont
  {B\"ottcher}, \citenamefont {Schmidt}, \citenamefont {Wenzel}, \citenamefont
  {Hertkorn}, \citenamefont {Guo}, \citenamefont {Langen},\ and\ \citenamefont
  {Pfau}}]{Bottcher2019}%
  \BibitemOpen
  \bibfield  {author} {\bibinfo {author} {\bibfnamefont {F.}~\bibnamefont
  {B\"ottcher}}, \bibinfo {author} {\bibfnamefont {J.-N.}\ \bibnamefont
  {Schmidt}}, \bibinfo {author} {\bibfnamefont {M.}~\bibnamefont {Wenzel}},
  \bibinfo {author} {\bibfnamefont {J.}~\bibnamefont {Hertkorn}}, \bibinfo
  {author} {\bibfnamefont {M.}~\bibnamefont {Guo}}, \bibinfo {author}
  {\bibfnamefont {T.}~\bibnamefont {Langen}},\ and\ \bibinfo {author}
  {\bibfnamefont {T.}~\bibnamefont {Pfau}},\ }\bibfield  {title} {\bibinfo
  {title} {Transient supersolid properties in an array of dipolar quantum
  droplets},\ }\href {https://doi.org/10.1103/PhysRevX.9.011051} {\bibfield
  {journal} {\bibinfo  {journal} {Phys. Rev. X}\ }\textbf {\bibinfo {volume}
  {9}},\ \bibinfo {pages} {011051} (\bibinfo {year} {2019})}\BibitemShut
  {NoStop}%
\bibitem [{\citenamefont {Chomaz}\ \emph {et~al.}(2019)\citenamefont {Chomaz},
  \citenamefont {Petter}, \citenamefont {Ilzh\"ofer}, \citenamefont {Natale},
  \citenamefont {Trautmann}, \citenamefont {Politi}, \citenamefont
  {Durastante}, \citenamefont {van Bijnen}, \citenamefont {Patscheider},
  \citenamefont {Sohmen}, \citenamefont {Mark},\ and\ \citenamefont
  {Ferlaino}}]{Chomaz2019}%
  \BibitemOpen
  \bibfield  {author} {\bibinfo {author} {\bibfnamefont {L.}~\bibnamefont
  {Chomaz}}, \bibinfo {author} {\bibfnamefont {D.}~\bibnamefont {Petter}},
  \bibinfo {author} {\bibfnamefont {P.}~\bibnamefont {Ilzh\"ofer}}, \bibinfo
  {author} {\bibfnamefont {G.}~\bibnamefont {Natale}}, \bibinfo {author}
  {\bibfnamefont {A.}~\bibnamefont {Trautmann}}, \bibinfo {author}
  {\bibfnamefont {C.}~\bibnamefont {Politi}}, \bibinfo {author} {\bibfnamefont
  {G.}~\bibnamefont {Durastante}}, \bibinfo {author} {\bibfnamefont {R.~M.~W.}\
  \bibnamefont {van Bijnen}}, \bibinfo {author} {\bibfnamefont
  {A.}~\bibnamefont {Patscheider}}, \bibinfo {author} {\bibfnamefont
  {M.}~\bibnamefont {Sohmen}}, \bibinfo {author} {\bibfnamefont {M.~J.}\
  \bibnamefont {Mark}},\ and\ \bibinfo {author} {\bibfnamefont
  {F.}~\bibnamefont {Ferlaino}},\ }\bibfield  {title} {\bibinfo {title}
  {Long-lived and transient supersolid behaviors in dipolar quantum gases},\
  }\href {https://doi.org/10.1103/PhysRevX.9.021012} {\bibfield  {journal}
  {\bibinfo  {journal} {Phys. Rev. X}\ }\textbf {\bibinfo {volume} {9}},\
  \bibinfo {pages} {021012} (\bibinfo {year} {2019})}\BibitemShut {NoStop}%
\end{thebibliography}

%

\end{document}